\documentclass[12pt]{article}
\usepackage{ae,aecompl}

\usepackage[T1]{fontenc}
\usepackage[utf8]{inputenc}
\usepackage{geometry}
\usepackage{array}
\usepackage{float}
\usepackage{enumitem}
\usepackage{varwidth}
\usepackage{amsmath}
\usepackage{amsthm}
\usepackage{amssymb}
\usepackage{stmaryrd}
\usepackage{graphicx}
\usepackage{rotfloat}
\usepackage{setspace}
\usepackage[authoryear]{natbib}
\usepackage[pdfusetitle,
 bookmarks=true,bookmarksnumbered=false,bookmarksopen=false,
 breaklinks=false,pdfborder={0 0 1},backref=false,colorlinks=false]
 {hyperref}

\makeatletter

\providecommand{\tabularnewline}{\\}
\newenvironment{cellvarwidth}[1][t]
    {\begin{varwidth}[#1]{\linewidth}}
    {\@finalstrut\@arstrutbox\end{varwidth}}
\floatstyle{ruled}
\newfloat{algorithm}{tbp}{loa}
\providecommand{\algorithmname}{Algorithm}
\floatname{algorithm}{\protect\algorithmname}

\theoremstyle{plain}
\newtheorem{assumption}{\protect\assumptionname}
\theoremstyle{remark}
\newtheorem{rem}{\protect\remarkname}
\theoremstyle{plain}
\newtheorem{prop}{\protect\propositionname}

\usepackage{setspace}
\usepackage{pdflscape}
\usepackage{datetime}

\DeclareMathOperator{\Var}{Var}

\date{}

\makeatother

\providecommand{\assumptionname}{Assumption}
\providecommand{\propositionname}{Proposition}
\providecommand{\remarkname}{Remark}

\begin{document}
\title{Measurement Error and Counterfactuals in \\ Quantitative Trade and Spatial Models}
\author{Bas Sanders, \textit{Harvard University and SEO Amsterdam Economics}\textbf{\thanks{E-mail: bas\_sanders@g.harvard.edu. I thank my advisors, Isaiah Andrews,
Pol Antràs, Anna Mikusheva and Jesse Shapiro, for their guidance and
generous support. I also thank Kevin Chen, Dave Donaldson, Tilman
Graff, Elhanan Helpman, Gabriel Kreindler, Marc Melitz, Ferdinando
Monte, Elie Tamer, Davide Viviano and Chris Walker for helpful discussions.
I am also grateful for comments from participants of the Harvard Graduate
Student Workshops in Econometrics and Trade, the 2024 UEA Summer School
and the 2025 North American Winter Meeting of the Econometric Society.}}}
\date{\monthname\ \number\year}
\maketitle
\begin{abstract}
Counterfactuals in quantitative trade and spatial models are functions
of the current state of the world and the model parameters. Common
practice treats the current state of the world as perfectly observed,
but there is good reason to believe that it is measured with error.
This paper provides tools for quantifying uncertainty about counterfactuals
when the current state of the world is measured with error. I recommend
an empirical Bayes approach to uncertainty quantification, and show
that it is both practical and theoretically justified. I apply the
proposed method to the settings in \citet{adao2017nonparametric}
and \citet{allen2022welfare} and find non-trivial uncertainty about
counterfactuals.
\end{abstract}

\section{Introduction}

Economists use quantitative trade and spatial models to evaluate counterfactual
scenarios. For instance, how do expenditure patterns across countries
adjust in response to the implementation of a trade agreement? How
are welfare levels affected when transportation infrastructure connecting
regions is improved? These counterfactual questions are typically
posed relative to an observed factual situation. This implies that
the estimand of interest depends directly on the realized data, rather
than on the underlying data-generating distribution—a departure from
standard statistical settings. 

The setting with data-dependent counterfactual estimands is further
complicated by the fact that data in quantitative trade and spatial
models are often measured with error \citep{goes2023reliability,linsi2023problem,teti2023missing}.
Unlike classical measurement error settings, where the estimand is
typically a parameter of the correctly measured population distribution,
here it is a functional of the realized data. To illustrate, consider
the canonical Armington model \citep{armington1969theory}, where
predicted welfare changes from hypothetical trade cost shocks can
be written as a function of baseline bilateral trade flows and the
trade elasticity \citep{arkolakis2012new}. The question I address
is how measurement error in the observed trade flows affects uncertainty
in the welfare predictions.

I develop an empirical Bayes framework for quantifying uncertainty
around counterfactual predictions. The approach requires specifying
both a measurement error model and a prior distribution over the latent
true data, up to a set of hyperparameters. These hyperparameters are
estimated from the observed data via an empirical Bayes step. Bayes’
rule then yields an estimated posterior distribution over the latent
data given the noisy observations. Given the structure of quantitative
trade and spatial models, this posterior induces a corresponding posterior
over counterfactual predictions. Uncertainty can then be summarized
by reporting an interval of posterior quantiles.

A natural accompanying point estimator for this procedure is the posterior
median, which is always guaranteed to be contained in the reported
interval. By contrast, the standard point estimator, which does not
account for measurement error, may lay outside the interval. The posterior
median answers the question: What does a Bayesian believe the counterfactual
prediction would have been in the absence of measurement error? While
this is a natural and intuitive question to ask, the answer necessarily
depends on the prior.

In settings where the observed data consist of non-negative dyadic
flows, I propose a default specification for the measurement error
model and prior that can be calibrated directly from the data, yielding
a widely applicable empirical Bayes approach. Specifically, I model
measurement error as log-normal and use a log-normal prior centered
on a structural gravity equation, with a point mass at zero to accommodate
zero flows. This setup is designed for ease of implementation and
is suitable for a wide class of quantitative trade and spatial models.

I consider two approaches to calibrating the hyperparameters in this
default specification. The first assumes constant measurement error
variances across flows and relies on researcher input or domain knowledge.
The second, applicable in the case of international trade, uses the
mirror trade dataset compiled by \citet{linsi2023problem}, which
reports bilateral trade flows as recorded by both exporters and importers.
I interpret these paired observations as two independent noisy measurements
of the true trade flow, enabling calibration of flow-specific measurement
error variances.

To illustrate the impact of incorporating measurement error into counterfactual
analysis, I revisit the applications in \citet{adao2017nonparametric}
and \citet{allen2022welfare}. In \citet{adao2017nonparametric},
which quantify the welfare effects of China’s accession to the WTO,
I model measurement error in baseline bilateral trade flows. I apply
the default empirical Bayes approach and construct uncertainty intervals
that account for measurement error in the estimated changes in China’s
welfare from 1996 to 2011. These intervals are substantially wider
than those reported in \citet{adao2017nonparametric}, which reflect
estimation uncertainty.

In the setting of \citet{allen2022welfare}, the counterfactual question
concerns which highway links in the United States yield the highest
return on investment and are therefore most promising for improvement.
I model measurement error in traffic flows and apply the default empirical
Bayes approach, calibrating the prior and measurement error model
using estimates from \citet{musunuru2019applications}. I compute
uncertainty intervals that account for measurement error for the three
links with the highest estimated returns. Although the intervals are
wide, the relative ranking of the top three links remains robust.

This paper contributes to a growing body of work aimed at improving
counterfactual analysis in quantitative trade and spatial models \citep{balistreri2008gravity,adao2017nonparametric,kehoe2017quantitative,adao2023putting,ansari2024quantifying,sanders2025new}.
The most closely related work is \citet{DingelTintelnot:2025}, which
studies calibration procedures in granular environments. That paper
considers models that presume a continuum of agents and shows that,
when data are limited, unit-level idiosyncrasies are absorbed into
the model parameters, leading to overfitting and poor out-of-sample
performance. My focus is on the complementary issue of uncertainty
quantification due to measurement error—an issue that persists even
in non-granular settings. \citet{DingelTintelnot:2025} recommends
replacing raw observed data with fitted values from a low-dimensional
model. I show how this recommendation can be nested into the proposed
Bayesian framework.

The remainder of the paper is organized as follows. Section \ref{sec:Counterfactuals_in_QTSMs}
introduces the setting and notation. Section \ref{sec:EBUQ} presents
the empirical Bayes framework for accounting for measurement error
in quantitative trade and spatial models. Section \ref{sec:Default_approach}
describes a widely applicable default approach. Section \ref{sec:Armington}
demonstrates the procedure in the context of the Armington model.
Section \ref{sec:Applications} applies the method to the trade setting
in \citet{adao2017nonparametric} and explores its use in the economic
geography framework of \citet{allen2022welfare}. Section \ref{sec:Conclusion}
concludes. 

\section{\protect\label{sec:Counterfactuals_in_QTSMs}Counterfactuals in Quantitative
Trade and Spatial Models}

This section introduces the notation and discusses the key assumption
that commonly underlies counterfactual analyses in quantitative trade
and spatial models. 

\subsection{Notation and Key Assumption}

To begin, consider a baseline setting where there is no measurement
error. Let $D\in\mathcal{D}\subseteq\mathbb{R}^{d_{D}}$ denote a
data vector drawn from distribution $\mathcal{P}_{D}$, and let $\theta\in\Theta\subseteq\mathbb{R}^{d_{\theta}}$
denote a structural parameter. Our objective is to compute a scalar
counterfactual quantity $\gamma\in\mathbb{R}$. The key assumption
that the counterfactual object of interest has to satisfy is: 
\begin{assumption}
\label{assu:Key_assumption}For a given counterfactual question and
fixed parameter value $\theta$, the counterfactual object $\gamma$
can be expressed as a function of the realized data $D$:
\begin{align}
\gamma & =g\left(D,\theta\right),\label{eq:g_k}
\end{align}
for some known function $g:\mathcal{D}\times\Theta\rightarrow\mathbb{R}$.
\end{assumption}
The exact functional form of $g$ depends on the specific quantitative
model that is considered. In Appendix \ref{sec:Finding_g} I discuss
Assumption \ref{assu:Key_assumption} for two leading classes of models,
namely invertible models and exact hat algebra models.

The main appeal of focusing on objects of the form in Assumption \ref{assu:Key_assumption}
is that it allows researchers to answer counterfactual questions posed
relative to a specific, observed factual situation. In quantitative
trade and spatial settings, such questions are often  at least as
relevant as those concerning average effects. For example, in a quantitative
model of international trade, the goal is typically to understand
what would happen to the world following a specific policy change,
rather than what would occur in a randomly drawn year under that policy.

Assumption \ref{assu:Key_assumption} implies that if the data $D$
are observed without error and the structural parameter $\theta$
is known, we can perfectly recover $\gamma$.\footnote{Indeed, by fixing $g$ I abstract away from model misspecification,
an important problem I engage with in future work.} This contrasts with standard econometric models, where the object
of interest is a function of the correctly measured distribution of
the data, rather than the actual observations. So the key distinction
with standard settings is:
\begin{equation}
\begin{cases}
\begin{array}{c}
\mathrm{standard\ setting:}\\
\mathrm{this\:paper:}
\end{array} & \begin{array}{c}
\gamma=g\left(\mathcal{P}_{D},\theta\right)\\
\gamma=g\left(D,\theta\right),\quad D\sim\mathcal{P}_{D}
\end{array}.\end{cases}\label{eq:Key_distinction}
\end{equation}
Importantly, this difference implies that it would not suffice to
be able to perfectly estimate the distribution $\mathcal{P}_{D}$.
Towards uncertainty quantification, we hence need to account for uncertainty
about the realized data themselves rather than their distribution.

\section{\protect\label{sec:EBUQ}Empirical Bayes Uncertainty Quantification}

This section introduces measurement error into quantitative trade
and spatial models. It outlines how to quantify the resulting uncertainty
for the counterfactual prediction of interest.

\subsection{Prior and Measurement Error Model}

Under Assumption \ref{assu:Key_assumption}, our object of interest
can be written as a function solely of the data realizations and the
structural parameter, which is convenient for answering relevant counterfactual
questions. However, the data realizations are economic variables which
are often measured with error. For instance, \citet{ortiz2018international}
and \citet{goes2023reliability} highlight that there are large discrepancies
between and within various data sources from trade and international
economics. Motivated by this, I assume that, instead of the true data
vector $D$, we observe a noisy version $\tilde{D}$.

For uncertainty quantification for the counterfactual prediction,
we will require the posterior distribution of the true data given
the noisy data. Towards that end, I introduce a model for the measurement
error and a prior distribution for the true underlying data,
\[
\begin{cases}
\begin{array}{c}
\mathrm{prior:}\\
\mathrm{measurement\:error:}
\end{array} & \begin{array}{c}
\pi^{\mathrm{prior}}\left(D|\vartheta\right)\\
\pi^{\mathrm{me}}\left(\tilde{D}|D,\vartheta\right)
\end{array}.\end{cases}
\]
Here, $\vartheta\in\mathbb{R}^{d_{\vartheta}}$ is a vector of unknown
hyperparameters. 

\subsection{Empirical Bayes and Posterior Distribution}

Given such a prior distribution and a measurement error model, we
can use an empirical Bayes approach to estimate the unknown hyperparameters.\footnote{Rather than estimating the parameters of the prior distribution for
the true underlying data, which corresponds to an empirical Bayes
approach, one could alternatively specify prior distributions for
these parameters, which corresponds to a hierarchical Bayes approach.} Formally, we have
\[
\tilde{\vartheta}=\underset{\vartheta}{\arg\max}\int\pi^{\mathrm{me}}\left(\tilde{D}|D,\vartheta\right)\pi^{\mathrm{prior}}\left(D|\vartheta\right)dD.
\]
Then, given the estimated hyperparameters $\tilde{\vartheta}$, we
can use Bayes' rule to find the estimated posterior distribution of
the true data given the noisy data,

\begin{equation}
\pi^{\mathrm{post}}\left(D|\tilde{D},\tilde{\vartheta}\right)=\frac{\pi^{\mathrm{me}}\left(\tilde{D}|D,\tilde{\vartheta}\right)\pi^{\mathrm{prior}}\left(D|\tilde{\vartheta}\right)}{\int\pi^{\mathrm{me}}\left(\tilde{D}|D,\tilde{\vartheta}\right)\pi^{\mathrm{prior}}\left(D|\tilde{\vartheta}\right)dD}.\label{eq:me_post}
\end{equation}
Using this estimated posterior, we can generate draws for the true
data given the noisy data.\footnote{Note that the measurement error distribution does not have to be mean
zero, so also allows for measurement error bias. Nevertheless, even
mean zero measurement error can result in bias in the counterfactual
prediction of interest. This is automatically taken into account by
the Bayesian approach when quantifying uncertainty. Furthermore, the
individual measurement error distributions can be arbitrarily correlated
in this general setup.} 

The Bayesian approach allows researchers to incorporate economic knowledge
through the prior. For example when considering measurement error
in non-negative flows between locations, one can fit a prior centered
on a gravity model, which I will do in Section \ref{sec:Default_approach}. 

\subsection{\protect\label{subsec:Quantifying_uncertainty_gamma}Quantifying
Uncertainty about $\gamma$}

The object of interest is a function of the true data and the structural
parameter. Going forward, I will assume the structural parameter is
known, an assumption I will discuss in more detail in Section \ref{subsec:Estimation_error}.
Then, under Assumption \ref{assu:Key_assumption} it follows that
we can obtain the estimated posterior for the counterfactual object
of interest, $\pi^{\mathrm{post}}\left(\gamma|\tilde{D},\tilde{\vartheta}\right)$.

Towards uncertainty quantification, we want to sample from this posterior
and report the relevant quantiles.\footnote{Note that counterfactual predictions are typically derived as functions
of the full system of counterfactual equilibrium variables. Thus,
whether the researcher is ultimately interested in a scalar outcome,
a relative comparison, or a global average, the mechanics of uncertainty
quantification—drawing from the posterior over the true data and solving
for equilibrium—remain the same.} The entire procedure is summarized in Algorithm \ref{alg:EB_UQ_k}.
\begin{algorithm}[h]
\caption{\protect\label{alg:EB_UQ_k}Uncertainty quantification about $\gamma=g\left(D,\theta\right)$}

\begin{enumerate}
\item Input: prior $\pi^{\mathrm{prior}}\left(D|\vartheta\right)$, measurement
error model $\pi^{\mathrm{me}}\left(\tilde{D}|D,\vartheta\right)$,
noisy data $\tilde{D}$, structural parameter $\theta$, number of
bootstrap draws $B$, coverage level $1-\alpha$ (choose $B$ and
$\alpha$ such that $\alpha/2\cdot B\in\mathbb{N}$).
\item Empirical Bayes estimation step: $\tilde{\vartheta}=\underset{\vartheta}{\arg\max}\int\pi^{\mathrm{me}}\left(\tilde{D}|D,\vartheta\right)\pi^{\mathrm{prior}}\left(D|\vartheta\right)dD.$
\item Construct estimated posterior: $\pi^{\mathrm{post}}\left(D|\tilde{D},\tilde{\vartheta}\right)\propto\pi^{\mathrm{me}}\left(\tilde{D}|D,\tilde{\vartheta}\right)\pi^{\mathrm{prior}}\left(D|\tilde{\vartheta}\right)$.
\item For $b=1,...,B$,
\begin{enumerate}
\item Draw $D_{b}\sim\pi^{\mathrm{post}}\left(D|\tilde{D},\tilde{\vartheta}\right)$.
\item Compute $\gamma_{b}=g\left(D_{b},\theta\right).$
\end{enumerate}
\item Sort $\left\{ \gamma_{b}\right\} _{b=1}^{B}$ to obtain $\left\{ \gamma^{(b)}\right\} _{b=1}^{B}$
with $\gamma^{(1)}\leq\gamma^{(2)}\leq...\leq\gamma^{(B)}$.
\item Report $\left[\gamma^{(\alpha/2\cdot B)},\gamma^{(\left(1-\alpha/2\right)\cdot B)}\right].$
\end{enumerate}
\end{algorithm}

A natural accompanying point estimator for the procedure in Algorithm
\ref{alg:EB_UQ_k} is the posterior median, which is always guaranteed
to be contained in the reported interval. By contrast, the standard
point estimator $g\left(\tilde{D},\theta\right)$, which does not
account for measurement error, may lay outside the interval. The posterior
median corresponds to an optimal estimate under the estimated posterior
and under absolute value loss from a decision-theoretic perspective
(see for example Proposition 2.5.5 in \citealp{robert2007bayesian}). 

Note that the posterior median does not guarantee bias correction
in the frequentist sense. It answers the question: What does a Bayesian
believe the counterfactual prediction would have been in the absence
of measurement error? While this is a natural and intuitive question
to ask, the answer necessarily depends on the prior. But when the
prior reflects well-established economic relationships—such as gravity
patterns in quantitative trade and spatial models—the posterior median
provides a principled estimate of the counterfactual prediction.

\subsection{Relation to the Literature}

\subsubsection{\protect\label{subsec:ME_literature}Relation to Measurement Error
Literature}

The literature on measurement error in nonlinear models is extensive,
as reviewed in \citet{hu2015microeconomic} and \citet{schennach2016recent},
and the most closely related strand of measurement error literature
is that on nonseparable error models \citep{matzkin2003nonparametric,chesher2003identification,hoderlein2007identification,matzkin2008identification,hu2008instrumental,schennach2012local,song2015estimating}.
However, these results do not apply to my setting. 

The key distinguishing feature of the setting in this paper is that
the object of interest $\gamma$ directly depends on the correctly
measured data, because the equality in Assumption \ref{assu:Key_assumption}
is an exact statement. This is convenient for answering counterfactual
questions, and arises because counterfactual questions are typically
posed relative to an observed factual situation. In contrast, in standard
econometric methods of measurement error, the object of interest is
a function of the correctly measured distribution of the data, $\mathcal{P}_{D}$,
rather than the actual realized observations, $D$. This leads to
the key distinction in Equation \eqref{eq:Key_distinction}. 

This difference is important because in my setting, it would not suffice
to be able to perfectly estimate the distribution $\mathcal{P}_{D}$.
For example in a quantitative model of international trade, to answer
counterfactual questions we need the realized trade flows, rather
than the trade flow distribution from which they are drawn. In contrast,
in standard econometric models of measurement error, knowing this
distribution would suffice, because the estimands are functionals
of the correctly measured distribution of the data. By virtue of that,
we need to account for uncertainty about the observations themselves
rather than their distribution.

\subsubsection{\protect\label{subsec:Dingel_and_Tintelnot}Relation to \citet{DingelTintelnot:2025}}

The most relevant paper in the literature on improving counterfactual
calculations in quantitative trade and spatial economics is \citet{DingelTintelnot:2025},
which studies calibration procedures in granular settings. In these
settings, individual idiosyncrasies do not wash out and can cause
overfitting and poor performance out-of-sample. To deal with this,
\citet{DingelTintelnot:2025} proposes to, instead of the observed
data, either use fitted values obtained using a low-dimensional model
or smooth the data using matrix approximation techniques. Both of
these approaches can be cast as special cases of the proposed procedure
in Algorithm \ref{alg:EB_UQ_k}, by choosing a specific prior. 

Specifically, the main recommendation is to use a low-dimensional
model and is called the covariates-based approach. \citet{DingelTintelnot:2025}
considers a quantitative spatial model with $L$ individuals. Let
$\ell_{ij}$ denote the share of people residing in location $i$
and working in location $j$. The covariates-based approach then interprets
the observed commuting shares $\left\{ \tilde{\ell}_{ij}\right\} $
as a finite sample from a continuum model. This results in the maximum
likelihood model
\begin{equation}
\left\{ \tilde{\ell}_{ij}\cdot L\right\} |\vartheta\sim\mathrm{Multinomial}\left(\left\{ h_{ij}\left(\vartheta\right)\right\} \right),\label{eq:DT_MLE}
\end{equation}
for $\vartheta$ a set of hyperparameters and $h_{ij}\left(\vartheta\right)$
a model function which I discuss further in Appendix \ref{subsec:Dingel_and_Tintelnot_details}. 

The covariates-based approach in \citet{DingelTintelnot:2025} first
finds a maximum likelihood estimator $\tilde{\vartheta}$ for $\vartheta$
using the model in Equation \eqref{eq:DT_MLE}. Next, focusing on
a specific counterfactual object of interest denoted by $\gamma=g\left(\left\{ \ell_{ij}\right\} ,\theta\right)$
for some known structural parameter $\theta$ and function $g$, the
approach recommends using the fitted values $\left\{ h_{ij}\left(\tilde{\vartheta}\right)\right\} $
instead of the observed shares $\left\{ \tilde{\ell}_{ij}\right\} $
to compute counterfactuals. That is, the main recommendation is to
use the estimate
\[
\tilde{\gamma}^{DT}=g\left(\left\{ h_{ij}\left(\tilde{\vartheta}\right)\right\} ,\theta\right)
\]
instead of $g\left(\left\{ \tilde{\ell}_{ij}\right\} ,\theta\right)$.

To see how the covariates-based approach from \citet{DingelTintelnot:2025}
is nested in my Bayesian framework, consider the following prior and
measurement error model,
\begin{equation}
\begin{cases}
\begin{array}{c}
\mathrm{prior:}\\
\mathrm{measurement\:error:}
\end{array} & \begin{array}{c}
\ell_{ij}|\vartheta\sim\delta_{h_{ij}\left(\vartheta\right)},\quad i,j=1,...,n\\
\left\{ \tilde{\ell}_{ij}\cdot L\right\} |\vartheta\sim\mathrm{Multinomial}\left(\left\{ \ell_{ij}\right\} \right)
\end{array},\end{cases}\label{eq:Dingel_and_Tintelnot}
\end{equation}
where $\delta_{h_{ij}\left(\vartheta\right)}$ denotes the Dirac mass
at $h_{ij}\left(\vartheta\right)$, implying a degenerate prior. The
empirical Bayes step then combines the prior and measurement error
model to find
\[
\left\{ \tilde{\ell}_{ij}\cdot L\right\} |\vartheta\sim\mathrm{Multinomial}\left(\left\{ h_{ij}\left(\vartheta\right)\right\} \right),
\]
which overlaps with the model in Equation \eqref{eq:DT_MLE}, and
uses maximum likelihood estimation to estimate $\vartheta$ by $\tilde{\vartheta}$.
This yields the estimated prior and measurement error model
\[
\begin{cases}
\begin{array}{c}
\mathrm{prior:}\\
\mathrm{measurement\:error:}
\end{array} & \begin{array}{c}
\ell_{ij}|\tilde{\vartheta}\sim\delta_{h_{ij}\left(\tilde{\vartheta}\right)},\quad i,j=1,...,n\\
\left\{ \tilde{\ell}_{ij}\cdot L\right\} |\tilde{\vartheta}\sim\mathrm{Multinomial}\left(\left\{ \ell_{ij}\right\} \right)
\end{array}.\end{cases}
\]
Using Bayes' rule we can then find the estimated posterior for the
true commuting shares, 
\begin{equation}
\ell_{ij}|\left\{ \tilde{\ell}_{ij}\right\} ,\tilde{\vartheta}\sim\delta_{h_{ij}\left(\tilde{\vartheta}\right)},\quad i,j=1,...,n.\label{eq:DT_posteriors}
\end{equation}
Note that after the empirical Bayes estimation step, since the estimated
prior is degenerate, no information is taken from the estimated measurement
error model. 

The estimated posterior distributions in Equation \eqref{eq:DT_posteriors}
translate to an estimated posterior for $\gamma$, 
\[
\pi^{\mathrm{post}}\left(\gamma|\left\{ \tilde{\ell}_{ij}\right\} ,\tilde{\vartheta}\right)=\delta_{g\left(\left\{ h_{ij}\left(\tilde{\vartheta}\right)\right\} ,\theta\right)}.
\]
Indeed, this posterior is a point mass at the counterfactual prediction
that uses the fitted values $\left\{ h_{ij}\left(\tilde{\vartheta}\right)\right\} $
as inputs. It follows that the covariates-based approach is a special
case of Algorithm \ref{alg:EB_UQ_k} by choosing the prior and measurement
error model as in Equation \eqref{eq:Dingel_and_Tintelnot}. 

Note that if we follow Algorithm \ref{alg:EB_UQ_k} exactly, then
in step 4a—where we draw from the posterior distributions in Equation
\eqref{eq:DT_posteriors}—we will obtain the same values in each bootstrap
iteration. As a result, the interval constructed in step 6 will collapse
to a single point. This outcome is unsurprising, as the procedure
in \citet{DingelTintelnot:2025} focuses on point estimation and correcting
overfitting bias, rather than on quantifying estimation uncertainty.
Accordingly, if the sole concern is overfitting bias, their method
provides an appropriate approach.

The second recommendation in \citet{DingelTintelnot:2025} is to replace
the observed data with a smoothed version using matrix approximation
techniques. I discuss this approach in Appendix \ref{subsec:Dingel_and_Tintelnot_details}.

\subsection{\protect\label{subsec:Estimation_error}Estimation Error}

The counterfactual prediction of interest will typically depend on
a structural parameter $\theta$. It is common in applied work to
plug in a fixed value for the structural parameter taken from the
literature or obtained through data-driven methods, thus ignoring
the uncertainty associated with the estimation process. An exception
is \citet{adao2017nonparametric}, which reports confidence sets for
the counterfactual predictions of interest that account for estimation
error. 

Towards accounting for estimation error for quantitative trade and
spatial models in the presence of measurement error, let $\tilde{\theta}$
denote the estimator of the estimand $\theta$. This estimand is usually
a function of the distribution of the data $\mathcal{P}_{D}$. This
implies that, to address measurement error affecting the structural
parameter, one can apply the frequentist measurement error techniques
discussed in Section \ref{subsec:ME_literature} to find a bias-corrected
estimate, though the resulting correction will not admit a Bayesian
interpretation.

Alternatively, in Appendix \ref{sec:Estimation_error_details} I outline
a fully Bayesian approach that also considers estimation error. Specifically,
I assume that the posterior distribution of the structural parameter
$\theta$ given the true data $D$ is approximately normal, which
is justified under regularity conditions that are closely related
to those required for frequentist asymptotic normality. We then have
two different posteriors,
\[
\begin{cases}
\begin{array}{c}
\mathrm{estimation\ error\ posterior:}\\
\mathrm{measurement\:error\ posterior:}
\end{array} & \begin{array}{c}
\pi^{\mathrm{post,ee}}\left(\theta|D\right)\\
\pi^{\mathrm{post,me}}\left(D|\tilde{D},\tilde{\vartheta}\right)
\end{array}.\end{cases}
\]
As in Section \ref{subsec:Quantifying_uncertainty_gamma}, a natural
point estimator for the structural parameter is the median of the
estimated posterior given the noisy data,

\[
\pi^{\mathrm{post}}\left(\theta|\tilde{D},\tilde{\vartheta}\right)=\int\pi^{\mathrm{post,ee}}\left(\theta|D\right)\pi^{\mathrm{post,me}}\left(D|\tilde{D},\tilde{\vartheta}\right)dD.
\]
Using Assumption \ref{assu:Key_assumption}, we can also find the
posterior $\pi^{\mathrm{post,ee}}\left(\gamma|D\right)$, and it follows
that a natural point estimator for the counterfactual prediction is
the median of the estimated posterior
\begin{align*}
\pi^{\mathrm{post}}\left(\gamma|\tilde{D},\tilde{\vartheta}\right) & =\int\pi^{\mathrm{post,ee}}\left(\gamma|D\right)\pi^{\mathrm{post,me}}\left(D|\tilde{D},\tilde{\vartheta}\right)dD.
\end{align*}
In Appendix \ref{sec:Estimation_error_details} I describe how to
sample from the estimated posteriors $\pi^{\mathrm{post}}\left(\theta|\tilde{D},\tilde{\vartheta}\right)$
and $\pi^{\mathrm{post}}\left(\gamma|\tilde{D},\tilde{\vartheta}\right)$.
There, I also outline how to quantify uncertainty while jointly accounting
for estimation error and measurement error in a natural way. 

It is important to understand that Bayesian estimators such as the
medians of $\pi^{\mathrm{post}}\left(\theta|\tilde{D},\tilde{\vartheta}\right)$
and $\pi^{\mathrm{post}}\left(\gamma|\tilde{D},\tilde{\vartheta}\right)$
need not satisfy frequentist properties such as consistency, even
when the prior is well-specified. I elaborate on this possibility
in Appendix \ref{sec:Estimation_error_details}, by showing frequentist
inconsistency of a structural estimator in a stylized example. 

However, for models satisfying Assumption \ref{assu:Key_assumption},
I am not aware of a frequentist framework that jointly incorporates
estimation error and measurement error within a unified procedure
that permits both point estimation and uncertainty quantification.
By contrast, the proposed Bayesian approach accommodates both sources
of uncertainty within a single coherent framework. Moreover, it nests
the case in which only measurement error is present: as estimation
error vanishes, the procedure naturally reduces to the framework that
accounts solely for measurement error. I therefore recommend the Bayesian
approach when both sources of uncertainty are relevant, and either
the Bayesian or frequentist approach when only estimation error is
of concern. In the latter case, the proposed framework can still serve
as a robustness check with respect to measurement error.

\section{\protect\label{sec:Default_approach}Widely Applicable Default Approach}

This section proposes a default empirical Bayes approach that can
be applied in many settings, as it is both economically reasonable
for many quantitative trade and spatial models and computationally
convenient. It also discusses the toolkit that accompanies the paper.

\subsection{Default Prior and Measurement Error Model}

In many applications, the support of the true data and the sampling
process naturally suggest both a measurement error model and a prior.
For example, when observing migration shares constructed from count
data, a multinomial measurement error model arises naturally from
sampling variation. Since the true shares are non-negative and must
sum to one, a Dirichlet prior provides a coherent and tractable specification. 

For settings where such natural choices are not available, this section
proposes a widely applicable default approach for quantifying uncertainty
about the counterfactual prediction of interest. This default approach
can be applied out-of-the-box to many quantitative trade and spatial
models, but can also be easily adapted to other settings. It recommends
default choices for the prior distribution and measurement error model,
and discusses how to calibrate both based on observed data.

Concretely, consider the setting where we can write $\gamma=g\left(\left\{ F_{ij}\right\} ,\theta\right)$,
for $\left\{ F_{ij}\right\} $ a set of non-negative flows between
locations. This setup is commonplace in quantitative trade and spatial
models \citep{costinot2014trade,redding2017quantitative,proost2019can}.
I assume that both the prior distributions on the true flows and the
measurement errors are mixtures of a point mass at zero and a log-normal
distribution, a so-called spike-and-slab distribution \citep{mitchell1988bayesian}.
The point mass at zero is necessary because in both trade and spatial
applications bilateral flows of zeros are common, particularly when
considering more granular data \citep{helpman2008estimating,DingelTintelnot:2025}.
This prior and measurement error model imply that the posterior distribution
of the true flows given the noisy flows will also be a spike-and-slab
distribution. This mixture model is fairly flexible, and conjugacy
ensures computational tractability. Furthermore, I assume that the
prior mean exhibits a gravity relationship, for which there is strong
empirical evidence \citep{head2014gravity,allen201813}.\footnote{One can easily enrich this gravity prior by adding other ``distance''
variables such as differences in income or productivity, or by adding
dummies that indicate similarity such as contiguity or a common language,
see for example \citet{silva2006log}. I experimented with this but
the results do not change much.} This is summarized in the following assumption: 
\begin{assumption}
\label{assu:Default_prior_me}We have
\[
\begin{cases}
\begin{array}{c}
\mathrm{true\ zeros:}\\
\mathrm{spurious\ zeros:}\\
\mathrm{prior:}\\
\\\mathrm{likelihood:}\\
\\\end{array} & \begin{array}{c}
P_{ij}\sim\mathrm{Bern}\left(p_{ij}\right)\\
B_{ij}\sim\mathrm{Bern}\left(b_{ij}\right)\\
F_{ij}\sim P_{ij}\cdot\delta_{0}+\left(1-P_{ij}\right)\cdot e^{\mathcal{N}\left(\mu_{ij},s_{ij}^{2}\right)}\\
\mu_{ij}=\beta\log\mathrm{dist}_{ij}+\alpha_{i}^{\mathrm{orig}}+\alpha_{j}^{\mathrm{dest}}\\
\tilde{F}_{ij}|F_{ij}\sim\delta_{0}\cdot\mathbb{I}\left\{ F_{ij}=0\right\} \\
+\left[B_{ij}\cdot\delta_{0}+\left(1-B_{ij}\right)\cdot e^{\mathcal{N}\left(\log F_{ij},\varsigma_{ij}^{2}\right)}\right]\cdot\mathbb{I}\left\{ F_{ij}>0\right\} 
\end{array},\end{cases}
\]
for $i,j=1,...,n$, where $\delta_{0}$ denotes the Dirac mass at
zero, $\mathrm{dist}_{ij}$ denotes the distance between locations
$i$ and $j$, $\alpha_{i}^{\mathrm{orig}}$ is an origin fixed effect
and $\alpha_{j}^{\mathrm{dest}}$ is a destination fixed effect.
\end{assumption}
The probability that a bilateral trade flow is truly zero is denoted
by $p_{ij}$, and a true zero flow is assumed to always result in
an observed zero.\footnote{Assumption \ref{assu:Default_prior_me} implies that both true and
spurious zeros occur randomly. Alternatively, one could think about
incorporating endogenous zeros using selection mechanisms such as
in \citet{helpman2008estimating}.} The probability of a spurious zero—that is, an observed zero despite
a non-zero underlying true flow—is denoted by $b_{ij}$. The prior
means and variances are denoted by $\left\{ \mu_{ij}\right\} $ and
$\left\{ s_{ij}^{2}\right\} $, respectively. The flow-specific measurement
error variances are denoted by $\left\{ \varsigma_{ij}^{2}\right\} $.

Gather the hyperparameters in $\vartheta=\left(\left\{ p_{ij}\right\} ,\left\{ b_{ij}\right\} ,\beta,\left\{ \alpha_{i}^{\mathrm{orig}}\right\} ,\left\{ \alpha_{i}^{\mathrm{dest}}\right\} ,\left\{ s_{ij}^{2}\right\} ,\left\{ \varsigma_{ij}^{2}\right\} \right)$.
It follows that the posterior distribution for the true flow between
location $i$ and $j$, $F_{ij}$, given its noisy version, $\tilde{F}_{ij}$
is given by
\begin{equation}
F_{ij}|\tilde{F}_{ij},\vartheta\sim\begin{cases}
\begin{array}{c}
Q_{ij}\cdot\delta_{0}+\left(1-Q_{ij}\right)\cdot e^{\mathcal{N}\left(\mu_{ij},s_{ij}^{2}\right)}\\
\exp\left\{ \mathcal{N}\left(\frac{s_{ij}^{2}}{s_{ij}^{2}+\varsigma_{ij}^{2}}\log\tilde{F}_{ij}+\frac{\varsigma_{ij}^{2}}{s_{ij}^{2}+\varsigma_{ij}^{2}}\mu_{ij},\left(\frac{1}{s_{ij}^{2}}+\frac{1}{\varsigma_{ij}^{2}}\right)^{-1}\right)\right\} 
\end{array} & \begin{array}{c}
\tilde{F}_{ij}=0\\
\tilde{F}_{ij}>0
\end{array},\end{cases}\label{eq:Posterior}
\end{equation}
for $i,j=1,...,n$, where $Q_{ij}\sim\mathrm{Bern}\left(\frac{p_{ij}}{p_{ij}+b_{ij}\left(1-p_{ij}\right)}\right)$.\footnote{Note that $F_{ij}$ are drawn independently across $ij$-pairs. As
a result, certain adding-up constraints may no longer hold. In principle,
this issue could be addressed by drawing from a constrained joint
distribution.} 

Conditional on being able to calibrate the parameters $\vartheta$—the
empirical Bayes estimation step—one can quantify uncertainty about
$\gamma$ by finding the interval as described in Algorithm \ref{alg:EB_UQ_k}.
Then, a default procedure for quantifying uncertainty about $\gamma$
is summarized in Algorithm \ref{alg:EB_UQ_k_Default_approach}.
\begin{algorithm}[h]
\caption{\protect\label{alg:EB_UQ_k_Default_approach}Uncertainty quantification
about $\gamma=g\left(\left\{ F_{ij}\right\} ,\theta\right)$}

\begin{enumerate}
\item Input: noisy flows $\left\{ \tilde{F}_{ij}\right\} $, structural
parameter $\theta$, number of bootstrap draws $B$, coverage level
$1-\alpha$ (choose $B$ and $\alpha$ such that $\alpha/2\cdot B\in\mathbb{N}$).
\item Empirical Bayes estimation step: calibrate $\vartheta$ as outlined
in Section \ref{subsec:Calibrating_vartheta} and denote the estimator
by $\tilde{\vartheta}$.
\item For $b=1,...,B$,
\begin{enumerate}
\item For $i,j=1,...,n$, draw $F_{ij,b}$ from the estimated posterior
distribution $\pi^{\mathrm{post}}\left(F_{ij}|\tilde{F}_{ij},\tilde{\vartheta}\right)$
as in Equation \eqref{eq:Posterior}.
\item Compute $\gamma_{b}=g_{}\left(\left\{ F_{ij,b}\right\} _{i,j=1}^{n},\theta\right)$.
\end{enumerate}
\item Sort $\left\{ \gamma_{b}\right\} _{b=1}^{B}$ to obtain $\left\{ \gamma^{(b)}\right\} _{b=1}^{B}$
with $\gamma^{(1)}\leq\gamma^{(2)}\leq...\leq\gamma^{(B)}$.
\item Report $\left[\gamma^{(\alpha/2\cdot B)},\gamma^{(\left(1-\alpha/2\right)\cdot B)}\right].$
\end{enumerate}
\end{algorithm}

\begin{rem}
\label{rem:Checking_normality}One can verify how reasonable the normality
assumption on the prior and measurement error model is by comparing
the histogram of the normalized residuals 
\[
\left\{ \frac{\log\tilde{F}_{ij}-\left\{ \tilde{\beta}\log\mathrm{dist}_{ij}+\tilde{\alpha}_{i}^{\mathrm{orig}}+\tilde{\alpha}_{j}^{\mathrm{dest}}\right\} }{\sqrt{\tilde{s}_{ij}^{2}+\tilde{\varsigma}_{ij}^{2}}}\right\} 
\]
with the probability density function of a standard normal distribution.
To further check the reasonableness of the gravity prior, we can look
at the adjusted R-squared of the gravity model and, following \citet{allen201813},
plot the log flows against the log distance for positive flows, after
partitioning out the origin and destination fixed effects. In Appendices
\ref{sec:Details_ACD} and \ref{sec:Details_AA} I perform both these
checks for my applications. 
\end{rem}
\begin{rem}
\label{rem:Sensitivity_analysis} One might be worried about misspecification
of the prior and measurement error model. For the normal-normal model,
we can use prior density-ratio classes to find worst-case bounds on
posterior quantiles over a neighborhood that contains distributions
that are not too far away from the assumed normal distribution for
the prior and measurement error model. It turns out that incorporating
uncertainty around the prior and measurement error model amounts to
reporting slightly wider quantiles. The details can be found in Appendix
\ref{sec:Misspecification_ME_prior}.
\end{rem}

\subsection{\protect\label{subsec:Calibrating_vartheta}Empirical Bayes Estimation
Step: Calibrating $\vartheta$}

The hyperparameters in $\vartheta$ need to be calibrated. I consider
two cases. 

\subsubsection{\protect\label{subsec:Calibrating_vartheta_baseline}Baseline Case
with Domain Knowledge}

In the baseline case I restrict the measurement error variances and
prior variances to be constant across flows so that $\varsigma_{ij}^{2}=\varsigma^{2}$
and $s_{ij}^{2}=s^{2}$ for all $i,j=1,...,n$. Furthermore, I require
knowledge of the common measurement error variance $\varsigma^{2}$
and of the Bernoulli parameters $\left\{ p_{ij}\right\} $ and $\left\{ b_{ij}\right\} $.\footnote{In the absence of a prior on the measurement error variance, one could
adopt a sensitivity analysis approach by plotting intervals while
varying the variance. Alternatively, one could determine the minimum
level of measurement error that would overturn the counterfactual
conclusion.} It then remains to estimate $\left(\beta,\left\{ \alpha_{i}^{\mathrm{orig}}\right\} ,\left\{ \alpha_{i}^{\mathrm{dest}}\right\} ,s^{2}\right)$.
Towards this, we can combine the equations in Assumption \ref{assu:Default_prior_me}
to find
\[
\log\tilde{F}_{ij}\sim\mathcal{N}\left(\beta\log\mathrm{dist}_{ij}+\alpha_{i}^{\mathrm{orig}}+\alpha_{j}^{\mathrm{dest}},s^{2}+\varsigma^{2}\right),\quad\tilde{F}_{ij}>0.
\]
Using maximum likelihood estimation, it follows that the prior mean
parameters can be estimated from the regression 
\[
\log\tilde{F}_{ij}=\beta\log\mathrm{dist}_{ij}+\alpha_{i}^{\mathrm{orig}}+\alpha_{j}^{\mathrm{dest}}+\phi_{ij},\quad\tilde{F}_{ij}>0,
\]
with $\phi_{ij}$ an error term. It follows that the estimated prior
means and variance are
\begin{align}
\tilde{\mu}_{ij} & =\tilde{\beta}\log\mathrm{dist}_{ij}+\tilde{\alpha}_{i}^{\mathrm{orig}}+\tilde{\alpha}_{j}^{\mathrm{dest}},\quad i,j=1,...,n\label{eq:prior_mean}\\
\tilde{s}^{2} & =\max\left\{ \widetilde{\Var}\left(\log\tilde{F}_{ij}-\tilde{\mu}_{ij}|\tilde{F}_{ij}>0\right)-\tilde{\varsigma}^{2},0\right\} .\label{eq:prior_variance}
\end{align}
Obtaining estimators for these prior means and variances is what \citet{walters2024empirical}
calls the deconvolution step.

\subsubsection{\protect\label{subsec:Calibrating_vartheta_mirror_trade}Mirror Trade
Data}

When the non-negative bilateral flows correspond to trade flows between
countries, I use the mirror trade dataset from \citet{linsi2023problem}
to calibrate $\vartheta$. This dataset has two estimates of each
bilateral trade flow, both as reported by the exporter and as by the
importer. \citet{linsi2023problem} shows that there are so-called
mirror discrepancies in bilateral trade flows between almost all countries.
This means that, for instance, while the value that Germany reports
it imported from France and the value that France reports it exported
to Germany should be the same, in practice they are often different.
I interpret this as observing two independent noisy observations per
time period for each bilateral trade flow. The key identifying assumptions
are that the flow-specific probabilities of true zeros, the flow-specific
probabilities of spurious zeros, and the flow-specific measurement
error variances are constant over time. 

The details for the calibration can be found in Appendix \ref{sec:Calibration_mirror_trade_data}.
I first calibrate the probabilities of true zeros $\left\{ p_{ij}\right\} $
and the probabilities of spurious zeros $\left\{ b_{ij}\right\} $
by noting that for each bilateral trade flow we can use the time variation
to identify the probabilities of observing a certain number of zeros.
I then leverage the model structure to calibrate the measurement error
variances $\left\{ \varsigma_{ij}^{2}\right\} $. Lastly, I calibrate
the prior parameters, using a similar approach as for the baseline
case with domain knowledge. To leverage country information and the
fact that importers and exporters can differ in their reliability,
I shrink the measurement error and prior variances using country-origin
and country-destination fixed effects.

\subsection{Toolkit}

Accompanying the paper, I provide an easy-to-use toolkit that consists
of three programs.\footnote{The toolkit is written in MATLAB and can be found on my website, https://sandersbas.github.io/.
A version in R is available upon request.} The first program implements the high-level approach in Algorithm
\ref{alg:EB_UQ_k}. It takes as inputs $\left(B,\theta,\tilde{D},\pi^{\mathrm{post}}\left(D|\tilde{D},\tilde{\vartheta}\right),g\right)$
and outputs posterior draws $\left\{ \gamma_{b}\right\} _{b=1}^{B}$.
The second program implements the default approach in Algorithm \ref{alg:EB_UQ_k_Default_approach}.
It takes as inputs $\left(B,\theta,\left\{ \tilde{F}_{ij}\right\} ,\tilde{\vartheta},g\right)$
and again outputs posterior draws $\left\{ \gamma_{b}\right\} _{b=1}^{B}$.
The third program, which can serve as an input to the second, uses
the mirror trade dataset of \citet{linsi2023problem} and allows the
researcher to choose countries and years for which they want to estimate
the hyperparameters of the prior and measurement error model. This
is summarized in Algorithm \ref{alg:Toolkit}. 
\begin{algorithm}[h]
\caption{\protect\label{alg:Toolkit}Toolkit}

\begin{enumerate}
\item Program 1: General algorithm.
\begin{itemize}
\item Input: number of draws $B$, structural parameter $\theta$, data
$\tilde{D}$, functions $\tilde{D}\mapsto D_{b}$, $\left(D,\theta\right)\mapsto\gamma$.
\item Output: posterior draws $\left\{ \gamma_{b}\right\} _{b=1}^{B}$.
\end{itemize}
\item Program 2: Default approach.
\begin{itemize}
\item Input: number of draws $B$, structural parameter $\theta$, noisy
flows $\left\{ \tilde{F}_{ij}\right\} $, estimated hyperparameters
$\tilde{\vartheta}$, function $\left(\left\{ F_{ij}\right\} ,\theta\right)\mapsto\gamma$.
\item Output: posterior draws $\left\{ \gamma_{b}\right\} _{b=1}^{B}$,
plot that compares histogram of the normalized residuals with the
probability density function of a standard normal distribution as
per Remark \ref{rem:Checking_normality}.
\end{itemize}
\item Program 3: Mirror trade data calibration.
\begin{itemize}
\item Input: countries $\mathcal{I}$, years to produce bootstrap draws
for $\mathcal{T}$, years to use for calibration $\mathcal{T}_{\mathrm{calibration}}$.
\item Output (can serve as input to Program 2): noisy flows $\left\{ \tilde{F}_{ij}\right\} $,
estimated hyperparameters $\tilde{\vartheta}$, adjusted R-squared
of the gravity model for the last year in $\mathcal{T}$, plot of
log flows against log distance for positive flows, after partitioning
out the origin and destination fixed effects as per Remark \ref{rem:Checking_normality}.
\end{itemize}
\end{enumerate}
\end{algorithm}

\section{\protect\label{sec:Armington}Prototypical Example: Armington Model}

This section illustrates the proposed procedure using the Armington
model \citep{armington1969theory}, a canonical workhorse model in
international trade, as outlined, for example, in \citet{costinot2014trade}.

\subsection{Model and Counterfactual Question of Interest}

Countries are indexed by $i,j=1,...,n$, and with CES preferences
and perfect competition, it follows that the relevant gravity equations
and budget constraints are:
\begin{align}
F_{ij} & =\frac{\left(\tau_{ij}Y_{i}\right)^{-\varepsilon}\chi_{ij}}{\sum_{k}\left(\tau_{kj}Y_{k}\right)^{-\varepsilon}\chi_{kj}}E_{j},\qquad i,j=1,...,n\label{eq:gravity}\\
E_{i} & =\left(1+\kappa_{i}\right)Y_{i},\qquad i=1,...,n.\label{eq:trade_deficit}
\end{align}
Here, $F_{ij}$ denotes the trade flow from country $i$ to $j$,
and $Y_{i}=\sum_{\ell=1}^{n}F_{i\ell}$, $E_{i}=\sum_{k=1}^{n}F_{ki}$
and $\kappa_{i}=\left(E_{i}-Y_{i}\right)/Y_{i}$ denote country $i$'s
total income, total expenditure and the ratio of the trade deficit
to income, respectively. Furthermore, $\tau_{ij}$ denotes the iceberg
trade cost between country $i$ and $j$, which means that in order
to sell one unit of a good in country $j$, country $i$ must ship
$\tau_{ij}\geq1$ units, with $\tau_{ii}=1$. Lastly, $\varepsilon>0$
is the trade elasticity and $\left\{ \chi_{ij}\right\} $ are idiosyncratic
preferences. 

Now, say we are interested in the counterfactual where we change the
trade costs $\left\{ \tau_{ij}\right\} $ proportionally by $\left\{ \tau_{ij}^{\mathrm{cf,prop}}\right\} $,
holding the trade elasticity $\varepsilon$, the idiosyncratic preferences
$\left\{ \chi_{ij}\right\} $ and the trade imbalance variables $\left\{ \kappa_{i}\right\} $
constant. In Appendix \ref{subsec:Derivation_Armington} I show that
we can then solve for the corresponding proportional changes in income,
$\left\{ Y_{i}^{\mathrm{cf,prop}}\right\} $, using
\[
Y_{i}^{\mathrm{cf,prop}}Y_{i}=\sum_{j}\frac{\left(\tau_{ij}^{\mathrm{cf,prop}}Y_{i}^{\mathrm{cf,prop}}\right)^{-\varepsilon}}{\sum_{k}\lambda_{kj}\left(\tau_{kj}^{\mathrm{cf,prop}}Y_{k}^{\mathrm{cf,prop}}\right)^{-\varepsilon}}\lambda_{ij}\left(1+\kappa_{j}\right)Y_{j}^{\mathrm{cf,prop}}Y_{j},\qquad i=1,...,n,
\]
where $\lambda_{ij}=F_{ij}/E_{j}$ denotes the expenditure share that
country $j$ spends on goods from country $i$. By Walras' Law, the
proportional changes in income are only pinned down up to a multiplicative
constant. Subsequently, following \citet{costinot2014trade}, we can
exactly solve for proportional changes in expenditure shares and welfare
(real consumption) levels:
\begin{align*}
\lambda_{ij}^{\mathrm{cf,prop}} & =\frac{\left(\tau_{ij}^{\mathrm{cf,prop}}Y_{i}^{\mathrm{cf,prop}}\right)^{-\varepsilon}}{\sum_{k}\lambda_{kj}\left(\tau_{kj}^{\mathrm{cf,prop}}Y_{k}^{\mathrm{cf,prop}}\right)^{-\varepsilon}},\qquad i,j=1,...,n\\
W_{i}^{\mathrm{cf,prop}} & =\left(\lambda_{ii}^{\mathrm{cf,prop}}\right)^{-1/\varepsilon},\qquad i=1,...,n.
\end{align*}
The income levels $\left\{ Y_{i}\right\} $, the expenditure shares
$\left\{ \lambda_{ij}\right\} $ and the trade deficit variables $\left\{ \kappa_{i}\right\} $
are all functions of the trade flows $\left\{ F_{ij}\right\} $, so
the relevant counterfactual mapping is
\[
\left\{ F_{ij}\right\} ,\left\{ \tau_{ij}^{\mathrm{cf,prop}}\right\} ,\varepsilon\mapsto\left\{ W_{i}^{\mathrm{cf,prop}}\right\} .
\]
It follows that for a given counterfactual question as described by
$\left\{ \tau_{ij}^{\mathrm{cf,prop}}\right\} $, we only require
knowledge of the baseline trade flows $\left\{ F_{ij}\right\} $ and
the trade elasticity $\varepsilon$. So we have:
\begin{align*}
D & =\left\{ F_{ij}\right\} \\
\theta & =\varepsilon.
\end{align*}
The specific counterfactual question I consider is a 10\% increase
in all bilateral trade costs between 76 countries, so that $\tau_{ij}^{\mathrm{cf,prop}}=1+0.1\cdot\mathbb{I}\left\{ i\neq j\right\} $
for $i,j=1,...,n$. I focus on the proportional percentage changes
in welfare in the Central African Republic, the Netherlands, Sweden
and the United States. It follows that, fixing $\left\{ \tau_{ij}^{\mathrm{cf,prop}}\right\} $,
we have
\begin{equation}
\gamma_{q}=100\cdot\left(W_{q}^{\mathrm{cf,prop}}-1\right)\equiv g_{q}\left(\left\{ F_{ij}\right\} ,\varepsilon\right),\label{eq:Armington_cf}
\end{equation}
for each $q\in\left\{ \mathrm{CAF},\mathrm{NLD},\mathrm{SWE},\mathrm{USA}\right\} $. 

\subsection{Measurement Error Model and Prior}

For the Armington model, I will consider measurement error in trade
flows $\left\{ F_{ij}\right\} $. Hence, instead of the true trade
flows we observe noisy trade flows $\left\{ \tilde{F}_{ij}\right\} $,
which in turn lead to noisy counterfactual predictions $\tilde{\gamma}_{q}$
for $q\in\left\{ \mathrm{CAF},\mathrm{NLD},\mathrm{SWE},\mathrm{USA}\right\} $.
Our goal is to quantify the uncertainty that arises from measurement
error, and report an accompanying point estimator that aims to answer
the question what a Bayesian believes the counterfactual predictions
would have been in the absence of measurement error.

If we specify a prior $\pi^{\mathrm{prior}}\left(\left\{ F_{ij}\right\} |\vartheta\right)$
and a measurement error model $\pi^{\mathrm{me}}\left(\left\{ \tilde{F}_{ij}\right\} |\left\{ F_{ij}\right\} ,\vartheta\right)$,
we can use empirical Bayes estimation and Bayes' rule to find the
estimated posterior $\pi^{\mathrm{post}}\left(\left\{ F_{ij}\right\} |\left\{ \tilde{F}_{ij}\right\} ,\tilde{\vartheta}\right)$.
The default approach from Section \ref{sec:Default_approach} can
be applied. For the empirical Bayes step, the calibration of $\vartheta$,
we can use the mirror trade data setting from Section \ref{subsec:Calibrating_vartheta_mirror_trade}.
So we can use the provided toolkit to obtain draws from $\pi^{\mathrm{post}}\left(\left\{ F_{ij}\right\} |\left\{ \tilde{F}_{ij}\right\} ,\tilde{\vartheta}\right)$.
I fix the trade elasticity to $\varepsilon=5$, a typical value in
the literature, which is also used in \citet{costinot2014trade}. 

\subsection{Results}

\begin{figure}[h]
\centering{}\includegraphics[scale=0.44]{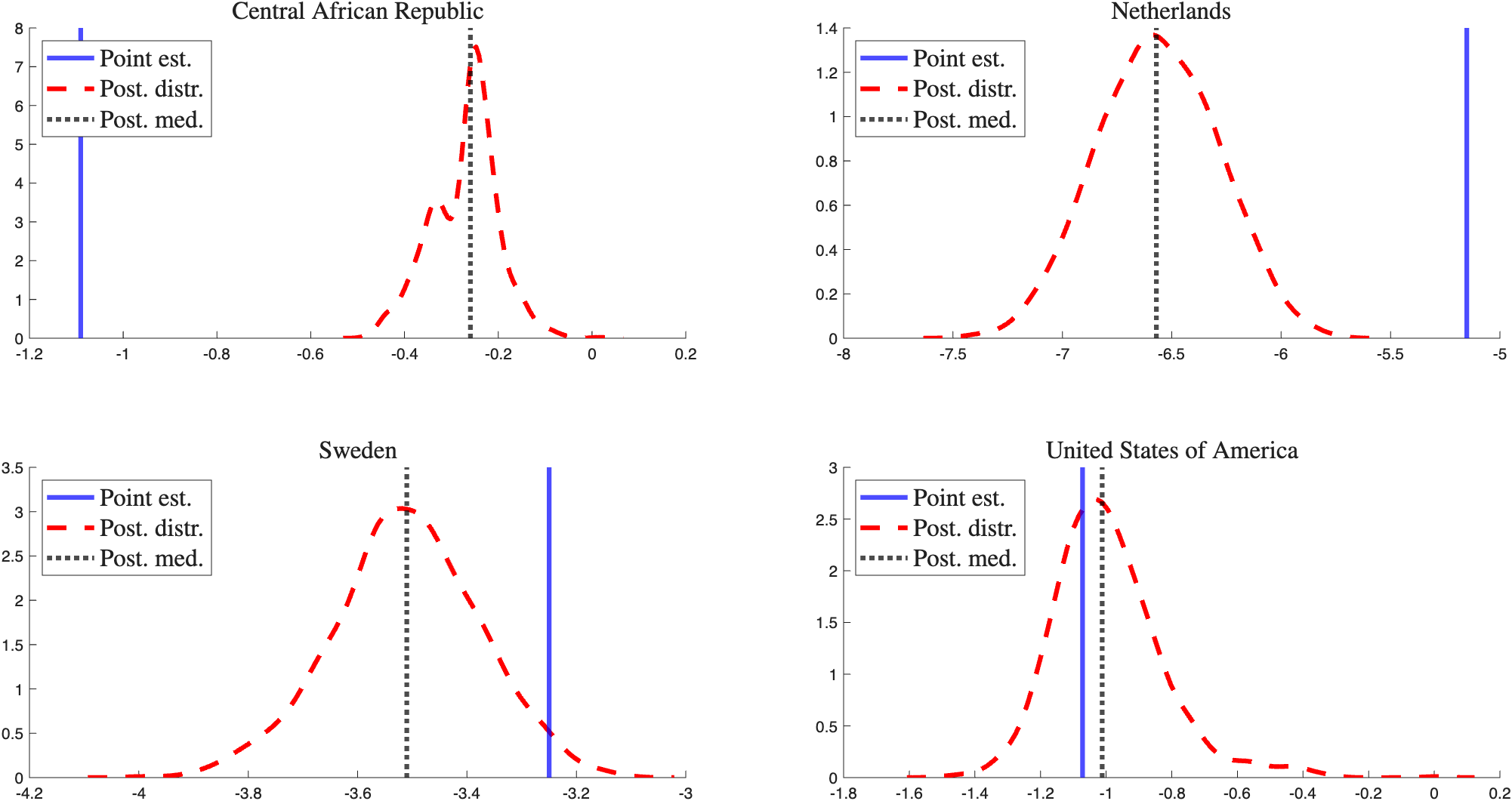}\caption{\protect\label{fig:Arm_gamma_ME}Uncertainty quantification for the
Armington model. The counterfactual object of interest is the percentage
change in welfare (real consumption) after a 10\% increase in all
bilateral trade costs. The solid blue line denotes the point estimate
$g\left(\left\{ \tilde{F}_{ij}\right\} ,5\right)$, the dashed red
line denotes the smoothed estimated posterior distribution $\pi^{\mathrm{post}}\left(g\left(\left\{ F_{ij}\right\} ,5\right)|\left\{ \tilde{F}_{ij}\right\} ,\tilde{\vartheta}\right)$,
and the dotted black line denotes the median of $\pi^{\mathrm{post}}\left(g\left(\left\{ F_{ij}\right\} ,5\right)|\left\{ \tilde{F}_{ij}\right\} ,\tilde{\vartheta}\right)$.}
\end{figure}

We can see the impact of measurement error in Table \ref{tab:Arm_gamma_ME}
and Figure \ref{fig:Arm_gamma_ME}. In Table \ref{tab:Arm_gamma_ME}
I compare the standard point estimates based on noisy flows and the
posterior median estimates, and report the intervals obtained using
Algorithm \ref{alg:EB_UQ_k_Default_approach}. In Figure \ref{fig:Arm_gamma_ME}
I plot the standard point estimates and the smoothed estimated posterior
distributions. 

We observe that for the Central African Republic and the Netherlands
there is a considerable difference between the point estimate and
the median of the posterior distribution, causing the point estimate
to lie outside the credible set. For Sweden and the United States
there is less of a discrepancy. These plots illustrate that the proposed
approach automatically incorporates bias induced by measurement error,
and that this bias can be both negative and positive. In Appendix
\ref{subsec:Other_countries}, I show the results for all 76 countries
in the sample. 
\begin{table}[h]
\begin{centering}
\begin{tabular}{|c|c|c|c|}
\hline 
 & \begin{cellvarwidth}[t]
\centering
Point estimate

$g\left(\left\{ \tilde{F}_{ij}\right\} ,5\right)$
\end{cellvarwidth} & \begin{cellvarwidth}[t]
\centering
Median of

$\pi^{\mathrm{post}}\left(g\left(\left\{ F_{ij}\right\} ,5\right)|\left\{ \tilde{F}_{ij}\right\} ,\tilde{\vartheta}\right)$
\end{cellvarwidth} & \begin{cellvarwidth}[t]
\centering
Interval accounting for

measurement error
\end{cellvarwidth}\tabularnewline
\hline 
\hline 
$\gamma_{\mathrm{CAF}}$ & -1.09 & -0.26 & {[}-0.42, -0.14{]}\tabularnewline
\hline 
$\gamma_{\mathrm{NLD}}$ & -5.15 & -6.57 & {[}-7.10, -6.04{]}\tabularnewline
\hline 
$\gamma_{\mathrm{SWE}}$ & -3.25 & -3.51 & {[}-3.79, -3.25{]}\tabularnewline
\hline 
$\gamma_{\mathrm{USA}}$ & -1.07 & -1.01 & {[}-1.27, -0.56{]}\tabularnewline
\hline 
\end{tabular}
\par\end{centering}
\caption{\protect\label{tab:Arm_gamma_ME}Uncertainty quantification for the
Armington model. The counterfactual object of interest is the percentage
change in welfare (real consumption) after a 10\% increase in all
bilateral trade costs. The intervals based on measurement error report
the 2.5th and 97.5th quantile of the estimated posterior distribution
$\pi^{\mathrm{post}}\left(g\left(\left\{ F_{ij}\right\} ,5\right)|\left\{ \tilde{F}_{ij}\right\} ,\tilde{\vartheta}\right)$.}
\end{table}

\section{\protect\label{sec:Applications}Applications}

In this section I discuss the applications in \citet{adao2017nonparametric}
and \citet{allen2022welfare}. The application in \citet{adao2017nonparametric}
extends the prototypical example to a panel-data setting and recovers
changes in trade costs using variation over time. The application
in \citet{allen2022welfare} illustrates how the proposed procedure
can be extended to an economic geography framework.

\subsection{\protect\label{subsec:ACD}Application 1: \citet{adao2017nonparametric}}

\subsubsection{Model and Counterfactual Question of Interest}

The empirical application of \citet{adao2017nonparametric} investigates
the effects of China joining the WTO, the so-called China shock. Specifically,
the authors examine what would have happened to China's welfare if
China's trade costs had stayed constant at their 1995 levels. They
consider $n$ countries and $T$ time periods. The exercise I am considering
is assessing the sensitivity of counterfactual predictions to measurement
error in bilateral trade flows. 

The counterfactual objects of interest is the change in China's welfare,
defined as the percentage change in income that the representative
agent in China would be indifferent about accepting instead of the
counterfactual change where China's trade costs are fixed at their
1995 levels. The details of the model can be found in Appendix \ref{subsec:Details_ACD_model}.\footnote{In \citet{adao2017nonparametric}, the authors consider two demand
systems: standard CES and “Mixed CES”. I focus on the standard CES
specification.} The key insight is that we can express the proportional change in
China's welfare in period $t$, denoted $W_{\mathrm{China},t}^{\mathrm{cf,prop}}$,
as a function of the full set of bilateral trade flows across periods,
$\left\{ F_{ij,t}\right\} $, and the trade elasticity, $\varepsilon$.
Implementing this mapping requires multiple years of data, since answering
the counterfactual question necessitates recovering the proportional
change in China’s trade costs between 1995 and year $t$. This change
is inferred from bilateral trade flows observed in those two years.
Hence, we can write
\begin{equation}
W_{\mathrm{China},t}^{\mathrm{cf,prop}}=g_{t}\left(\left\{ F_{ij,t}\right\} ,\varepsilon\right),\label{eq:ACD_g}
\end{equation}
for $t=1,...,T$ and known functions $g_{t}:\mathbb{R}_{+}^{Tn\left(n-1\right)}\times\mathbb{R}_{++}\rightarrow\mathbb{R}$.
Then, conditional on a prior distribution for the true bilateral flows
$\left\{ F_{ij,t}\right\} $ and a measurement error model, we can
quantify uncertainty for $\left\{ W_{\mathrm{China},t}^{\mathrm{cf,prop}}\right\} $. 

\subsubsection{Measurement Error Model and Prior}

The default approach from Section \ref{sec:Default_approach} can
be applied. For the empirical Bayes step, the calibration of $\vartheta$,
we can use the mirror trade data setting from Section \ref{subsec:Calibrating_vartheta_mirror_trade}.
Since there are no zero flows in this application, the estimated posterior
of interest is 
\[
F_{ij,t}|\tilde{F}_{ij,t}\sim\exp\left\{ \mathcal{N}\left(\frac{\mathring{s}_{ij}^{2}}{\mathring{s}_{ij}^{2}+\mathring{\varsigma}_{ij}^{2}}\log\left(\tilde{F}_{ij,t}\right)+\frac{\mathring{\varsigma}_{ij}^{2}}{\mathring{s}_{ij}^{2}+\mathring{\varsigma}_{ij}^{2}}\tilde{\mu}_{ij,t},\left(\frac{1}{\mathring{s}_{ij}^{2}}+\frac{1}{\mathring{\varsigma}_{ij}^{2}}\right)^{-1}\right)\right\} ,
\]
where $\left\{ \mathring{s}_{ij}^{2}\right\} $, $\left\{ \mathring{\varsigma}_{ij}^{2}\right\} $,
$\left\{ \tilde{F}_{ij,t}\right\} $ and $\left\{ \tilde{\mu}_{ij,t}\right\} $
are all defined in Appendix \ref{sec:Calibration_mirror_trade_data}.

\subsubsection{Results}

\begin{figure}[h]
\centering{}\includegraphics[scale=0.44]{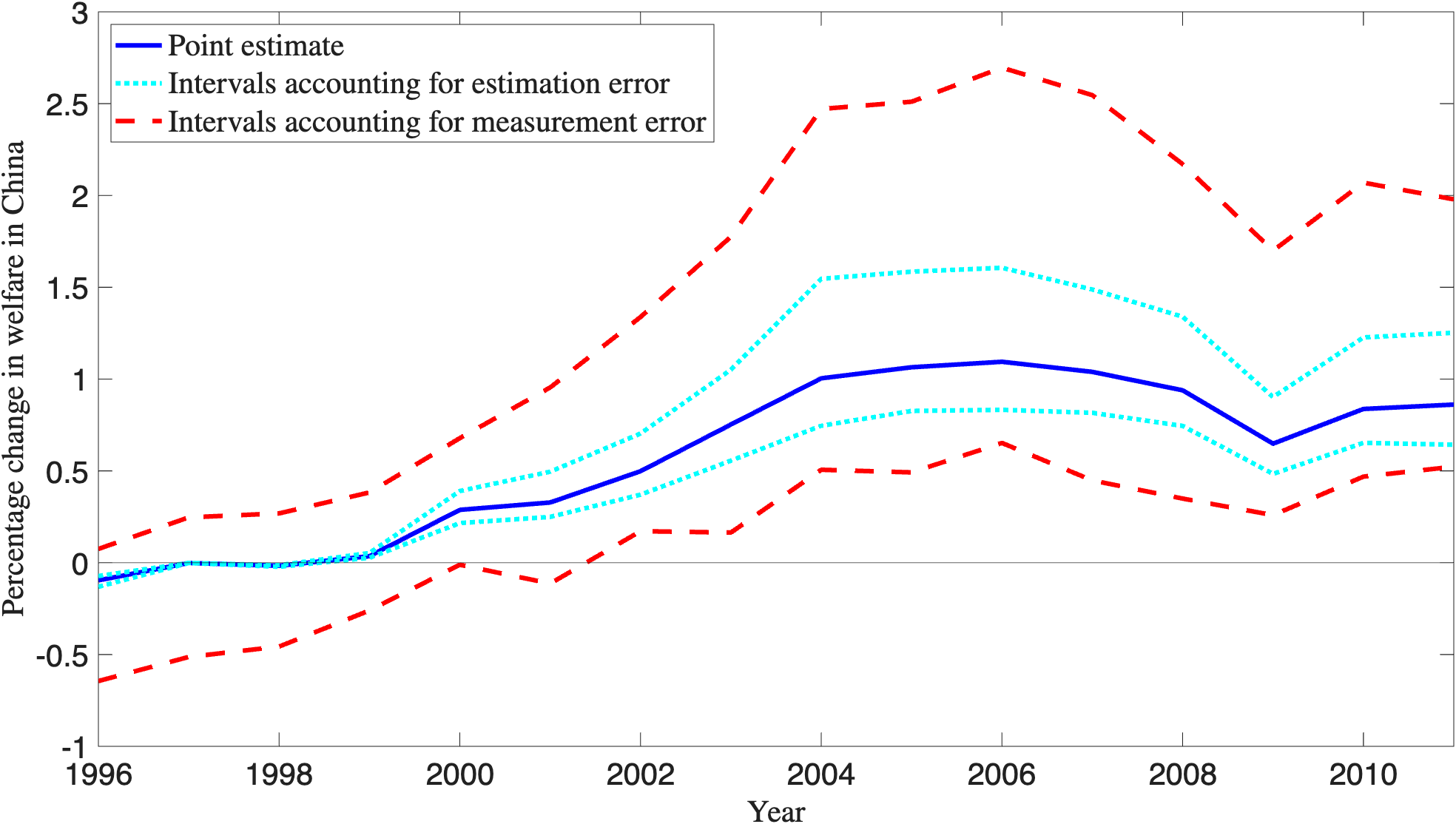}\caption{\protect\label{fig:ACD}Uncertainty quantification for heteroskedastic
normal shocks to $\left\{ \log F_{ij,t}\right\} $ for the percentage
change in China's welfare due to the China shock. The solid blue line
is the estimate as reported in \citet{adao2017nonparametric}, the
dotted light-blue lines denote the intervals accounting for estimation
error as reported in \citet{adao2017nonparametric}, and the dashed
red lines denote the intervals based on the estimated posterior distributions
$\pi^{\mathrm{post}}\left(g_{t}\left(\left\{ F_{ij,t}\right\} ,\varepsilon\right)|\left\{ \tilde{F}_{ij,t}\right\} ,\tilde{\vartheta}\right)$
for $t=1,...,T$. }
\end{figure}

Having obtained a posterior distribution for the true trade flows
given the noisy trade flows, we can now quantify uncertainty about
the counterfactual predictions of interest. In Figure \ref{fig:ACD},
I reproduce Figure 3 of \citet{adao2017nonparametric}, which plots
the percentage change in China's welfare as a result of the China
shock for each year in the period 1996-2011, and include two $95\%$
intervals. 

The first region only considers estimation error and hence assumes
the data are perfectly measured. It is constructed using code provided
by the authors, and samples from the normal distribution with mean
and variance equal to the GMM estimator for the trade elasticity $\varepsilon$
and its sampling variance, respectively. The resulting intervals are
small for the period before the year 2000, and then slowly become
wider. These are the intervals reported in \citet{adao2017nonparametric}.

The second region considers only measurement error and no estimation
error in $\varepsilon$. The resulting intervals are considerably
wider than the intervals based on estimation error, especially in
the first few years. In Appendix \ref{subsec:Details_ACD_supplementary_analyses}
I provide additional discussion and analyses. 

\subsection{Application 2: \citet{allen2022welfare}}

\subsubsection{Model and Counterfactual Question of Interest}

The empirical application in \citet{allen2022welfare} aims to estimate
the returns on investment for all highway segments of the US Interstate
Highway network. The authors do so by introducing an economic geography
model and calculating what happens to welfare after a $1\%$ improvement
to all highway links. Combining these counterfactual welfare changes
with how many lane-miles must be added in order to achieve the $1\%$
improvement, they find the highway segments with the greatest return
on investment.

This exercise only requires data on incomes and traffic flows of the
$n$ locations and knowledge of four structural model parameters.
The details of the model can be found in Appendix \ref{subsec:Details_AA_model},
but the key relation is the one that maps the average annual daily
traffic (AADT) flows $\left\{ F_{ij}\right\} $ to the counterfactual
return on investments $\left\{ R_{k\ell}^{\mathrm{cf}}\right\} $,
which is 
\[
R_{k\ell}^{\mathrm{cf}}=g_{k\ell}\left(\left\{ F_{ij}\right\} ,\theta\right)
\]
for known functions $g_{k\ell}:\mathbb{R}_{+}^{n\left(n-1\right)}\times\Theta\rightarrow\mathbb{R}$
for $k,\ell=1,...,n$ locations in the US Interstate Highway network. 

\subsubsection{Measurement Error Model and Prior}

For this application we can again apply the default approach from
Section \ref{sec:Default_approach}. For the empirical Bayes step
we can use the baseline case from Section \ref{subsec:Calibrating_vartheta_baseline}.
There are no zeros so we only have to provide an estimate of the measurement
error variance $\varsigma^{2}$. \citet{musunuru2019applications}
estimates that the measurement error variance of the logarithm of
the average annual daily traffic (AADT) flows, which is exactly the
data that \citet{allen2022welfare} uses, is between 0.05 and 0.20.
To obtain a lower bound on uncertainty, I will use a uniform measurement
error variance of $0.05$.

With $\tilde{\varsigma}^{2}=0.05$, I use Equation \eqref{eq:prior_variance}
to find a prior variance of $\tilde{s}^{2}=0.101$. This results in
the following estimated posterior distribution for the true traffic
flow between country $i$ and $j$, $F_{ij}$, given its noisy version
$\tilde{F}_{ij}$, for $i,j=1,...,n$:
\[
F_{ij}|\tilde{F}_{ij}\sim\mathrm{exp}\left\{ \mathcal{N}\left(0.669\cdot\log\tilde{F}_{ij}+0.331\cdot\tilde{\mu}_{ij},0.033\right)\right\} ,
\]
where $\tilde{\mu}_{ij}$ is defined in Equation \eqref{eq:prior_mean}. 

\subsubsection{Results}

The counterfactual question of interest is which links have the highest
return on investment, and the authors of \citet{allen2022welfare}
report the top ten links. For exposition, I will focus my analysis
on the three best performing links. Table \ref{tab:AA2022_table_bay}
shows the $95\%$ intervals for the top three links based on Algorithm
\ref{alg:EB_UQ_k_Default_approach}. 
\begin{table}[h]
\begin{centering}
\begin{tabular}{|c|c|c|}
\hline 
 & Point estimate & \begin{cellvarwidth}[m]
\centering
Interval accounting for

measurement error
\end{cellvarwidth}\tabularnewline
\hline 
\hline 
Link 1 & 10.43 & {[}8.69, 14.15{]}\tabularnewline
\hline 
Link 2 & 9.54 & {[}7.31, 10.83{]}\tabularnewline
\hline 
Link 3 & 7.31 & {[}6.78, 8.18{]}\tabularnewline
\hline 
\end{tabular}
\par\end{centering}
\caption{\protect\label{tab:AA2022_table_bay}Uncertainty quantification for
the return on investment for the three links from \citet{allen2022welfare}
with the highest return on investment. Returns on investment are reported
as annualized decimal returns (10.43 means an annual 1043\% return).
Link 1 is Kingsport-Bristol (TN-VA) to Johnson City (TN), link 2 is
Greensboro-High Point (NC) to Winston-Salem (NC) and link 3 is Rochester
(NY) to Batavia (NY). The intervals based on measurement error report
the 2.5th and 97.5th quantile of the estimated posterior distributions. }
\end{table}

From a policy perspective it is of interest whether the ranking between
these links can change due to measurement error. Therefore, Table
\ref{tab:AA2022_table_differences_bay} shows the $95\%$ intervals
for the difference between link 1 and link 2, and the difference between
link 2 and link 3.\footnote{This simple exercise is intended purely for exposition. For a more
formal treatment of inference on ranks, see \citet{mogstad2024inference}.} It follows that the rankings are generally robust against measurement
error.\textbf{ }Additional discussion and analyses can be found in
Appendices \ref{subsec:Details_AA_calibration} and \ref{subsec:Details_AA_supplementary_analyses}.
\begin{table}[h]
\begin{centering}
\begin{tabular}{|c|c|c|}
\hline 
 & Point estimate & \begin{cellvarwidth}[m]
\centering
Interval accounting for

measurement error
\end{cellvarwidth}\tabularnewline
\hline 
\hline 
Link 1-Link 2 & 0.89 & {[}0.38, 5.39{]}\tabularnewline
\hline 
Link 2-Link 3 & 2.23 & {[}-0.05, 3.27{]}\tabularnewline
\hline 
\end{tabular}
\par\end{centering}
\caption{\protect\label{tab:AA2022_table_differences_bay}Uncertainty quantification
for the differences in return on investment between the three links
from \citet{allen2022welfare} with the highest return on investment.
Returns on investment are reported as annualized decimal returns (0.89
means an annual 89\% return). Link 1 is Kingsport-Bristol (TN-VA)
to Johnson City (TN), link 2 is Greensboro-High Point (NC) to Winston-Salem
(NC) and link 3 is Rochester (NY) to Batavia (NY). The intervals based
on measurement error report the 2.5th and 97.5th quantile of the estimated
posterior distributions.}
\end{table}

\section{\protect\label{sec:Conclusion}Conclusion}

This paper develops an econometric framework for quantifying the impact
of measurement error in a broad class of quantitative trade and spatial
models. Unlike standard econometric models of measurement error, the
counterfactual estimand in these models depends directly on the realized
data rather than on the underlying distribution. I adopt an empirical
Bayes approach to characterize uncertainty in counterfactual predictions
and propose a default specification that can be easily implemented
across a range of applications. Applying the framework to the settings
in \citet{adao2017nonparametric} and \citet{allen2022welfare}, I
find substantial uncertainty surrounding key economic outcomes. These
results underscore the need to account for measurement error when
using quantitative models to guide policy decisions.

\bibliographystyle{econ-econometrica}
\bibliography{CSiEM}
\newpage{}

\appendix
\begin{center}
{\huge Appendix}{\huge\par}
\par\end{center}

\section{\protect\label{sec:Finding_g}Characterizing $g$ in Two Leading
Classes of Models}

Assumption \ref{assu:Key_assumption} requires that, for a fixed counterfactual
question and a fixed parameter value $\theta$, the counterfactual
object of interest can be written as $\gamma=g\left(D,\theta\right)$,
where $D$ collects the baseline inputs taken from the observed economy,
and $\theta$ collects model parameters that are held fixed across
baseline and counterfactual economies. This appendix discusses how
to construct $g$ in two broad classes of quantitative models, namely
invertible models and exact hat algebra models. 

\subsection{Invertible Models}

Following \citet{redding2017quantitative}, a model is \textit{invertible}
if, given $\theta$, the equilibrium conditions imply a one-to-one
mapping between the baseline observables $D$ and a set of latent
equilibrium objects $X\in\mathcal{X}\subseteq\mathbb{R}^{d_{X}}$,
often called \textit{fundamentals}. Formally, there exist maps $X=\mathcal{J}\left(D,\theta\right)$
and $D=\mathcal{E}\left(X,\theta\right)$, that are inverses of each
other (up to standard normalizations such as a price level or numeraire).
In quantitative trade and spatial models, $X$ often includes objects
such as trade costs and productivity levels. For example, in the Armington
model in Section \ref{sec:Armington}, the fundamentals are $X=\left(\left\{ \tau_{ij}\right\} ,\left\{ \chi_{ij}\right\} ,\left\{ \kappa_{i}\right\} \right)$. 

Now consider a counterfactual experiment that specifies \textit{proportional}
changes to the fundamentals $X$. Denote these proportional changes
by $X^{\mathrm{cf,prop}}\in\mathbb{R}^{d_{X}}$. For example in the
Armington model in Section \ref{sec:Armington}, the proportional
changes are $X^{\mathrm{cf,prop}}=\left(\left\{ 1+0.1\cdot\mathbb{I}\left\{ i\neq j\right\} \right\} ,\left\{ 1\right\} ,\left\{ 1\right\} \right)$.
The goal is to compute the corresponding proportional changes to the
observed data, $D^{\mathrm{cf,prop}}\in\mathbb{R}^{d_{D}}$, and then
the scalar counterfactual object $\gamma$. The computational steps
are:
\begin{enumerate}
\item ``Invert'' the model to recover the levels of the fundamentals $X$:
\[
X=\mathcal{J}\left(D,\theta\right).
\]
\item Solve for counterfactual levels of the data $D\varodot D^{\mathrm{cf,prop}}$
and convert to proportional changes:
\[
D\varodot D^{\mathrm{cf,prop}}=\mathcal{E}\left(X\varodot X^{\mathrm{cf,prop}},\theta\right)\quad\Rightarrow\quad D^{\mathrm{cf,prop}}=\mathcal{E}\left(X\varodot X^{\mathrm{cf,prop}},\theta\right)\oslash D,
\]
where $\varodot$ and $\oslash$ denote element-wise multiplication
and division, respectively.
\item Compute the counterfactual object of interest:
\[
\gamma=\Gamma\left(D,D^{\mathrm{cf,prop}},\theta\right).
\]
\end{enumerate}
Composing these steps yields a function of $\left(D,\theta\right)$
alone:
\[
\gamma=\Gamma\left(D,\mathcal{E}\left(\mathcal{J}\left(D,\theta\right)\varodot X^{\mathrm{cf,prop}},\theta\right)\oslash D,\theta\right)\equiv g\left(D,\theta\right).
\]

\subsection{Exact Hat Algebra Models}

Exact hat algebra models \citep{costinot2014trade} are models in
which counterfactuals can be computed directly in proportional changes
(``hats'') from baseline observables. In such models, there exist
a mapping, given $\theta$, from the proportional changes in fundamentals
and the observed data to the corresponding changes in the observed
data,
\[
D^{\mathrm{cf,prop}}=\mathcal{H}\left(X^{\mathrm{cf,prop}},D,\theta\right).
\]
This mapping holds holds directly, without the intermediary step of
backing out the levels of the fundamentals $X$. It follows that in
this case
\[
\gamma=\Gamma\left(D,\mathcal{H}\left(X^{\mathrm{cf,prop}},D,\theta\right),\theta\right)\equiv g\left(D,\theta\right).
\]
The Armington model presented in Section \ref{sec:Armington} is an
example of an exact hat algebra model. Since both invertible models
and exact hat algebra models satisfy Assumption \ref{assu:Key_assumption},
they are essentially equivalent for the purposes of uncertainty quantification.

\section{\protect\label{subsec:Dingel_and_Tintelnot_details}Details for Relation
to \citet{DingelTintelnot:2025}}

\subsection{Details for Covariates-Based Approach}

The specific form for $h_{ij}\left(\vartheta\right)$ in Equation
\eqref{eq:Dingel_and_Tintelnot} is
\[
h_{ij}\left(\vartheta\right)=\frac{w_{j}^{\varepsilon}\left(r_{i}^{\eta}\bar{\delta}_{ij}\right)^{-\varepsilon}}{\sum_{s,t}w_{t}^{\varepsilon}\left(r_{s}^{\eta}\bar{\delta}_{st}\right)^{-\varepsilon}},
\]
where $w_{j}$ denotes the wage in location $j$, $r_{i}$ denotes
the land rent in location $i$, $\bar{\delta}_{ij}$ denotes the time
component of commuting cost, $\varepsilon$ denotes the commuting
elasticity, and $\eta$ denotes a Cobb-Douglas preference parameter.
Since $\left\{ \bar{\delta}_{ij}\right\} $ are known, it follows
that $\vartheta=\left(\left\{ \alpha_{i}^{\mathrm{orig}}\right\} ,\left\{ \alpha_{j}^{\mathrm{dest}}\right\} ,\varepsilon\right)=\left(\left\{ r_{i}^{-\eta\varepsilon}\right\} ,\left\{ w_{j}^{\varepsilon}\right\} ,\varepsilon\right)$.

\subsection{Using Matrix Approximation Techniques}

In the truncated singular value decomposition approach in \citet{DingelTintelnot:2025},
the recommendation is to use an approximated matrix instead of the
matrix with noisy flows $\left\{ \tilde{\ell}_{ij}\right\} $. To
see this approach can be nested in my Bayesian framework by, defining
$\tilde{\mathcal{L}}\equiv\left\{ \tilde{\ell}_{ij}\right\} $ and
$\mathcal{L}\equiv\left\{ \ell_{ij}\right\} $, consider the following
prior and measurement error:
\begin{equation}
\begin{cases}
\begin{array}{c}
\mathrm{prior:}\\
\mathrm{measurement\:error:}
\end{array} & \begin{array}{c}
\pi^{\mathrm{prior}}\left(\mathcal{L}|A\right)=\delta_{A},\quad A\in\mathcal{A}=\left\{ B:\mathrm{rank}\left(B\right)\leq\tau\right\} \\
\tilde{\mathcal{L}}=\mathcal{L}+\xi,\quad\xi_{ij}\overset{\mathrm{iid}}{\sim}\mathcal{N}\left(0,1\right)
\end{array}.\end{cases}\label{eq:Dingel_and_Tintelnot_SVD}
\end{equation}
In this case, the empirical Bayes step solves:
\begin{align*}
\tilde{A} & =\underset{A\in\mathcal{A}}{\arg\max}\int\exp\left(-\frac{1}{2}\sum_{i,j}\left(\tilde{\mathcal{L}}_{ij}-\mathcal{L}_{ij}\right)^{2}\right)\delta_{A}\left(\mathcal{L}\right)d\mathcal{L}\\
 & =\underset{A\in\mathcal{A}}{\arg\max}\exp\left(-\left\Vert \tilde{\mathcal{L}}-A\right\Vert _{F}^{2}\right),
\end{align*}
where $\left\Vert \cdot\right\Vert _{F}$ denotes the Frobenius norm.
This maximization problem is equivalent to projecting the noisy flows
onto the space of matrices that have a rank no larger than $\tau$,
which, by the Eckart–Young–Mirsky theorem, is solved by the truncated
singular value decomposition. This yields the estimated prior and
measurement error model
\[
\begin{cases}
\begin{array}{c}
\mathrm{prior:}\\
\mathrm{measurement\:error:}
\end{array} & \begin{array}{c}
\pi^{\mathrm{prior}}\left(\mathcal{L}|\tilde{A}\right)=\delta_{\tilde{A}},\quad A\in\mathcal{A}=\left\{ B:\mathrm{rank}\left(B\right)\leq\tau\right\} \\
\tilde{\mathcal{L}}=\mathcal{L}+\xi,\quad\xi_{ij}\overset{\mathrm{iid}}{\sim}\mathcal{N}\left(0,1\right)
\end{array}.\end{cases}
\]
Using Bayes' rule we can then find the estimated posterior for the
true flows
\[
\pi^{\mathrm{post}}\left(\mathcal{L}|\tilde{\mathcal{L}},\tilde{A}\right)=\delta_{\tilde{A}}.
\]
So the posterior is a point mass at the counterfactual prediction
that uses the approximated matrix $\tilde{A}$. It follows that the
truncated singular value decomposition approach is a special case
of Algorithm \ref{alg:EB_UQ_k} by choosing the prior and measurement
error model as in Equation \eqref{eq:Dingel_and_Tintelnot_SVD}.

\section{\protect\label{sec:Estimation_error_details}Details for Estimation
Error}

The counterfactual prediction of interest will typically depend on
a structural parameter $\theta$. It is common in applied work to
plug in a fixed value for the structural parameter taken from the
literature or obtained through data-driven methods, thus ignoring
the uncertainty associated with the estimation process. I will discuss
two different approaches to dealing with estimation error.

\subsection{Frequentist Approach to Dealing with Estimation Error}

Let $\tilde{\theta}$ denote the estimator of the estimand $\theta$.
This estimand is usually a function of the distribution of the data
$\mathcal{P}_{D}$. This implies that, to address measurement error
affecting the structural parameter, one can apply the frequentist
measurement error techniques discussed in Section \ref{subsec:ME_literature}
to find a bias-corrected estimate, though the resulting correction
will not admit a Bayesian interpretation. For example, one may use
repeated measurements or instrumental variables to construct consistent
estimators.

\subsection{Bayesian Approach to Dealing with Estimation Error}

Alternatively, one can take a Bayesian or quasi-Bayesian approach
and assume that the posterior or quasi-posterior distribution of the
true structural parameter $\theta$ given the data $D$ is approximately
normal.\footnote{Formally, this normality could follow from assumptions on the underlying
data generating process such that a Bernstein-von Mises type result
holds \citep{van2000asymptotic}. In that case the influence of the
prior distribution $\pi\left(\theta\right)$ becomes negligible and
the posterior distribution approximately equals a normal distribution
centered at the maximum likelihood estimator. In \citet{sanders2025new}
I engage further with structural estimation in quantitative trade
and spatial models.} Specifically, we have the estimation error posterior (with superscript
``post,ee'')
\begin{equation}
\pi^{\mathrm{post,ee}}\left(\theta|D\right)\approx\mathcal{N}\left(\tilde{\theta}\left(D\right),\tilde{\Sigma}\left(D\right)\right),\label{eq:pi_EE}
\end{equation}
where $\tilde{\Sigma}\left(D\right)$ is a consistent estimator of
the sampling variance of $\tilde{\theta}\left(D\right)$.\footnote{This notation nests the scenario where we use an estimator from another
study that used different data. In that case $\theta$ is independent
from $D$ and we would write $\pi^{\mathrm{post,ee}}\left(\theta|D\right)\approx\mathcal{N}\left(\tilde{\theta},\tilde{\Sigma}\right)$.
Furthermore, in the case where $\theta$ is known to be non-negative,
one can use a log-normal distribution here. } 

We can then generate draws from the posterior distribution of $\theta$
given $D$.\footnote{Note that this assumption is on the structural parameter $\theta$,
and not on the fundamentals as discussed in Appendix \ref{sec:Finding_g}.
One could additionally use a degenerate posterior $\pi^{\mathrm{post,ee}}\left(X|D\right)=\delta_{X}$
on these fundamentals, since they are parameters that are linked deterministically
to the data $D$, and hence there is no estimation error.} For each of these draws, we can calculate the corresponding value
of the counterfactual object of interest using the relationship $\gamma=g\left(D,\theta\right)$.
This allows us to find the posterior distribution of $\gamma$ given
the true data, $\pi^{\mathrm{post,ee}}\left(\gamma|D\right)$.

\subsubsection{Combining Measurement Error and Estimation Error}

The object of interest is a function of the true data and the structural
parameter. It follows that we must consider estimation error, the
direct effect of mismeasurement, and the indirect effect of mismeasurement
through the estimation procedure. Our goal is to quantify uncertainty
about $\gamma$ when we observe $\tilde{D}$ by accounting for these
various sources of uncertainty. 

Recall that we have obtained two different posteriors. The first one
is the posterior distribution of $\gamma$ given the true data, $\pi^{\mathrm{post,ee}}\left(\gamma|D\right)$,
which incorporates estimation error. The second one is the posterior
of the true data given the noisy data, $\pi^{\mathrm{post,me}}\left(D|\tilde{D},\tilde{\vartheta}\right)$,
which incorporates measurement error (with superscript ``post,me'').
We can combine these two posteriors uncertainty quantification and
point estimation for $\gamma$.\footnote{Note that one could in principle use a single prior $\pi$ on the
underlying data generating process to handle both estimation error
and measurement error. I instead combine two simple priors to separately
handle estimation error and measurement error, since this leads to
highly tractable procedures, albeit at the cost of complicating the
Bayesian interpretation of resulting intervals.} 

For uncertainty quantification, I recommend to report an interval
$\mathcal{C}$ to which, in posterior expectation over $D$, the posterior
$\pi^{\mathrm{post,ee}}\left(\gamma|D\right)$ assigns probability
$1-\alpha$:
\[
\mathbb{E}_{\pi^{\mathrm{post,me}}}\left[Pr_{\pi^{\mathrm{post,ee}}}\left\{ \gamma\in\mathcal{C}|D\right\} |\tilde{D},\tilde{\vartheta}\right]\geq1-\alpha.
\]
In practice, given $\tilde{D}$ one would generate draws from $\pi^{\mathrm{post,me}}\left(D|\tilde{D},\tilde{\vartheta}\right)$,
and for each of these draws obtain a corresponding draw from $\pi^{\mathrm{post,ee}}\left(\gamma|D\right)$.\footnote{If an estimator from another study is used, then $\pi^{\mathrm{post,ee}}\left(\theta|D\right)\approx\mathcal{N}\left(\tilde{\theta},\tilde{\Sigma}\right)$.
In that case, we can draw $\theta_{b}$ and $D_{b}$ separately, which
makes the algorithm much faster.} Then, one would report the $\alpha/2$ and $1-\alpha/2$ quantiles
of this second set of draws.\footnote{When obtaining draws from $\pi^{\mathrm{post,me}}\left(D|\tilde{D},\tilde{\vartheta}\right)$
is computationally expensive, it could help improve computational
speed to take multiple draws of $\theta_{b}$ for the same $D_{b}$.} This is summarized in Algorithm \ref{alg:EB_UQ_k_ME_EE}.
\begin{algorithm}[h]
\caption{\protect\label{alg:EB_UQ_k_ME_EE}Uncertainty quantification about
$\gamma=g\left(D,\theta\right)$}

\begin{enumerate}
\item Input: prior $\pi^{\mathrm{prior}}\left(D|\vartheta\right)$, measurement
error model $\pi^{\mathrm{me}}\left(\tilde{D}|D,\vartheta\right)$,
quasi-posterior $\pi^{\mathrm{post,ee}}\left(\theta|D\right)$, noisy
data $\tilde{D}$, number of bootstrap draws $B$, coverage level
$1-\alpha$ (choose $B$ and $\alpha$ such that $\alpha/2\cdot B\in\mathbb{N}$).
\item Empirical Bayes estimation step: $\tilde{\vartheta}=\underset{\vartheta}{\arg\max}\int\pi^{\mathrm{me}}\left(\tilde{D}|D,\vartheta\right)\pi^{\mathrm{prior}}\left(D|\vartheta\right)dD.$
\item Construct estimated posterior: $\pi^{\mathrm{post,me}}\left(D|\tilde{D},\tilde{\vartheta}\right)\propto\pi^{\mathrm{me}}\left(\tilde{D}|D,\tilde{\vartheta}\right)\pi^{\mathrm{prior}}\left(D|\tilde{\vartheta}\right)$.
\item For $b=1,...,B$,
\begin{enumerate}
\item Draw $D_{b}\sim\pi^{\mathrm{post,me}}\left(D|\tilde{D},\tilde{\vartheta}\right)$.
\item Draw $\theta_{b}\sim\pi^{\mathrm{post,ee}}\left(\theta|D_{b}\right)$.
\item Compute $\gamma_{b}=g\left(D_{b},\theta_{b}\right).$
\end{enumerate}
\item Sort $\left\{ \gamma_{b}\right\} _{b=1}^{B}$ to obtain $\left\{ \gamma^{(b)}\right\} _{b=1}^{B}$
with $\gamma^{(1)}\leq\gamma^{(2)}\leq...\leq\gamma^{(B)}$.
\item Report $\left[\gamma^{(\alpha/2\cdot B)},\gamma^{(\left(1-\alpha/2\right)\cdot B)}\right].$
\end{enumerate}
\end{algorithm}

As in Section \ref{subsec:Quantifying_uncertainty_gamma}, a natural
point estimator for the structural parameter is the median of the
estimated posterior of the structural parameter given the noisy data,
\[
\pi^{\mathrm{post}}\left(\theta|\tilde{D},\tilde{\vartheta}\right)=\int\pi^{\mathrm{post,ee}}\left(\theta|D\right)\pi^{\mathrm{post,me}}\left(D|\tilde{D},\tilde{\vartheta}\right)dD,
\]
and a natural point estimator for the counterfactual prediction is
the median of the estimated posterior,
\begin{align*}
\pi^{\mathrm{post}}\left(\gamma|\tilde{D},\tilde{\vartheta}\right) & =\int\pi^{\mathrm{post,ee}}\left(\gamma|D\right)\pi^{\mathrm{post,me}}\left(D|\tilde{D},\tilde{\vartheta}\right)dD.
\end{align*}
These posterior medians can be calculated using the draws obtained
in steps 4b and 4c in Algorithm \ref{alg:EB_UQ_k_ME_EE}, respectively. 

I illustrate this procedure in Appendix \ref{subsec:Details_ACD_supplementary_analyses}
by applying it to the application in \citet{adao2017nonparametric}.

\subsubsection{Frequentist Consistency}

A caveat of the proposed Bayesian approach to dealing with estimation
error is that it will not guarantee frequentist consistency of the
estimator. To see this, consider the stylized model in which the independent
variable in a simple regression is measured with error:
\[
\begin{cases}
\begin{array}{c}
\mathrm{regression:}\\
\mathrm{measurement\ error:}\\
\mathrm{prior:}
\end{array} & \begin{array}{c}
\theta=\frac{\mathrm{Cov}\left(Y_{i},X_{i}\right)}{\mathrm{Var}\left(X_{i}\right)}\\
\tilde{X}_{i}=X_{i}+\varepsilon_{i}\\
X_{i}=\beta Z_{i}+\nu_{i}
\end{array},\end{cases}
\]
where $\varepsilon_{i}$, $Z_{i}$ and $\nu_{i}$ are mean-zero normal
random variables. In this setting, the IV estimator $\frac{\mathrm{\widehat{Cov}}\left(Y_{i},Z_{i}\right)}{\mathrm{\widehat{Cov}}\left(\tilde{X}_{i},Z_{i}\right)}$
is consistent provided an exogeneity and relevance condition hold. 

Alternatively, using the proposed empirical Bayesian approach, we
would estimate the hyperparameter $\beta$ by noting that
\[
\tilde{X}_{i}=\beta Z_{i}+\nu_{i}+\varepsilon_{i},
\]
and we would use the empirical Bayes estimator 
\[
\tilde{\beta}=\frac{\mathrm{\widehat{Cov}}\left(\tilde{X_{i}},Z_{i}\right)}{\mathrm{\widehat{Var}}\left(Z_{i}\right)}\overset{p}{\rightarrow}\beta+\frac{\mathrm{\mathrm{Cov}}\left(\nu_{i}+\varepsilon_{i},Z_{i}\right)}{\mathrm{\mathrm{Var}}\left(Z_{i}\right)}.
\]
Under this setup, for the posterior weight $w_{\mathrm{po}}\in\left[0,1\right]$
computed as in Section \ref{sec:Default_approach}, draws from the
posterior $X_{i}^{*}\sim\pi^{\mathrm{post}}\left(X_{i}|\tilde{X}_{i},Z_{i},\tilde{\beta}\right)$
can be represented as
\begin{align*}
X_{i}^{*} & =w_{\mathrm{po}}\cdot\tilde{X}_{i}+\left(1-w_{\mathrm{po}}\right)\cdot\tilde{\beta}Z_{i}\\
 & \overset{p}{\rightarrow}w_{\mathrm{po}}\cdot\tilde{X}_{i}+\left(1-w_{\mathrm{po}}\right)\cdot\beta Z_{i}+\left(1-w_{\mathrm{po}}\right)\frac{\mathrm{\mathrm{Cov}}\left(\nu_{i}+\varepsilon_{i},Z_{i}\right)}{\mathrm{\mathrm{Var}}\left(Z_{i}\right)}Z_{i}\\
 & =X_{i}+\underbrace{w_{\mathrm{po}}\cdot\varepsilon_{i}-\left(1-w_{\mathrm{po}}\right)\cdot\nu_{i}+\left(1-w_{\mathrm{po}}\right)\frac{\mathrm{\mathrm{Cov}}\left(\nu_{i}+\varepsilon_{i},Z_{i}\right)}{\mathrm{\mathrm{Var}}\left(Z_{i}\right)}Z_{i}}_{\equiv p_{i}}.
\end{align*}
Here, $p_{i}$ captures the deviation of the posterior draw from the
truth, which has mean zero but non-zero variance. The probability
limit of the regression coefficient based on posterior draws then
is
\[
\frac{\mathrm{Cov}\left(Y_{i},X_{i}+p_{i}\right)}{\mathrm{Var}\left(X_{i}+p_{i}\right)},
\]
which does not generally equal $\theta=\frac{\mathrm{Cov}\left(Y_{i},X_{i}\right)}{\mathrm{Var}\left(X_{i}\right)}$.

\section{\protect\label{sec:Details_ACD}Details for Application \citet{adao2017nonparametric}}

\subsection{\protect\label{subsec:Details_ACD_model}Model Details}

In the empirical application of \citet{adao2017nonparametric}, the
authors investigate the effects of China joining the WTO, the so-called
China shock. Going forward, $Q_{i,t}$ denotes the factor endowment
of country $i$ in period $t$, $\tau_{ij,t}$ denotes the trade cost
between country $i$ and $j$ in period $t$, $\lambda_{ij,t}$ denotes
the expenditure share from country $i$ in country $j$ in period
$t$, $Y_{i,t}$ denotes the income of country $i$ in period $t$,
and $P_{i,t}$ denotes the factor price of country $i$ in period
$t$. Furthermore, $\rho_{i,t}$ denotes the difference between aggregated
gross expenditure and gross production in country $i$ in period $t$,
which is assumed to stay constant for different counterfactuals. Lastly,
$\varepsilon$ denotes the trade elasticity and $\chi_{i}\left(\cdot\right)$
denotes the factor demand system of country $i$. 

In \citet{adao2017nonparametric}, two demand systems are considered,
normal CES and ``Mixed CES''. I will focus on normal CES, so that
\[
\lambda_{ij,t}=\chi_{i}\left(\left\{ \delta_{ij,t}\right\} \right)=\frac{\exp\left\{ \delta_{ij,t}\right\} }{1+\sum_{\ell>1}\exp\left\{ \delta_{i\ell,t}\right\} },
\]
for $\delta_{ij,t}$ some transformation of factor prices. The function
$\chi_{i}^{-1}\left(\cdot\right)$ then maps the observed expenditures
shares to values of this transformation. The structural parameter
$\varepsilon$ is estimated by assuming a model on the unobserved
trade costs $\left\{ \tau_{ij,t}\right\} $, and is estimated using
GMM with as an input the expenditure shares $\left\{ \lambda_{ij,t}\right\} $.

The counterfactual question of interest is what the change in China's
welfare is due to joining the WTO. This question is modeled by choosing
the counterfactual proportional changes in trade costs, $\left\{ \tau_{ij,t}^{\mathrm{cf,prop}}\right\} $,
such that Chinese trade costs are brought back to their 1995 levels:
\begin{align*}
\tau_{ij,t}^{\mathrm{cf,prop}} & =\frac{\tau_{ij,95}}{\tau_{ij,t}},\quad\mathrm{if}\:i\:\mathrm{or}\:j\:\mathrm{is}\:\mathrm{China,}\\
\tau_{ij,t}^{\mathrm{cf,prop}} & =1,\quad\mathrm{otherwise.}
\end{align*}
Welfare is then defined as the percentage change in income that the
representative agent in China would be indifferent about accepting
instead of the counterfactual change in trade costs from $\left\{ \tau_{ij,t}\right\} $
to $\left\{ \tau_{ij,t}^{\mathrm{cf,prop}}\tau_{ij,t}\right\} $.
These proportional changes in China's welfare $\left\{ W_{\mathrm{China},t}^{\mathrm{cf,prop}}\right\} $
can be obtained from first solving for $\left\{ P_{i,t}^{\mathrm{cf,prop}}\right\} $
using the system of equations 
\[
\sum_{j}\frac{\exp\left\{ \chi_{i}^{-1}\left(\left\{ \lambda_{ij,t}\right\} \right)-\varepsilon\log\left(P_{i,t}^{\mathrm{cf,prop}}\tau_{ij,t}^{\mathrm{cf,prop}}\right)\right\} }{1+\sum_{\ell>1}\exp\left\{ \chi_{\ell}^{-1}\left(\left\{ \lambda_{ij,t}\right\} \right)-\varepsilon\log\left(P_{\ell,t}^{\mathrm{cf,prop}}\tau_{\ell j,t}^{\mathrm{cf,prop}}\right)\right\} }\left\{ P_{j,t}^{\mathrm{cf,prop}}Y_{j,t}+\rho_{j,t}\right\} =P_{i,t}^{\mathrm{cf,prop}}Y_{i,t},
\]
and then using 
\[
W_{i,t}^{\mathrm{cf,prop}}=100\cdot\left(P_{i,t}^{\mathrm{cf,prop}}\frac{\sum_{\ell}\left[\chi_{\ell}^{-1}\left(\left\{ \lambda_{ij,t}\right\} \right)\right]^{-\varepsilon}}{\sum_{\ell}\left[P_{\ell,t}^{\mathrm{cf,prop}}\tau_{i\ell,t}^{\mathrm{cf,prop}}\left(\left\{ \lambda_{ij,t}\right\} \right)\right]^{-\varepsilon}}-1\right).
\]

\subsection{\protect\label{subsec:Details_ACD_calibration}Calibration Procedure
and Computational Details}

The default approach from Section \ref{sec:Default_approach} can
be applied. For the empirical Bayes step, the calibration of $\vartheta$,
we can use the mirror trade data setting from Section \ref{subsec:Calibrating_vartheta_mirror_trade}. 

In preprocessing the mirror trade dataset from \citet{linsi2023problem}
I made some additional assumptions. Firstly, I only consider data
from the period that is considered in \citet{adao2017nonparametric}.
Secondly, I only consider trade flows between countries that the authors
of that paper consider. This amounts to aggregating Belgium and Luxembourg,
and Estonia and Latvia. All the remaining countries I aggregate to
``Rest of World''. Thirdly, when only one of the mirror trade flows
is reported, I interpret this as zero measurement error by setting
the unknown mirror trade flow equal to the observed one. Relatedly,
when both mirror trade flows are not reported, I interpret this as
there being no trade, and when one trade flow is zero and the other
is substantially larger than zero, I set the zero trade flow equal
to the non-zero one. Lastly, I follow \citet{adao2017nonparametric}
by setting zero trade flows to 0.0025 (million USD). There are however
only a handful of zeros due to the aggregation into ``Rest of World''.

When estimating the prior distribution of the true underlying trade
flows, I use the distance dataset from \citet{mayer2011notes}. For
the distance between countries and the ``Rest of World'', I take
the average of the distances to all other countries that are considered
in \citet{adao2017nonparametric}. 

An important consideration is that there is a substantial difference
between the trade flows used in \citet{adao2017nonparametric}, which
come from the World Input Output Dataset (WIOD), and the mirror trade
flows from \citet{linsi2023problem}, which are based on the IMF Direction
of Trade Statistics dataset. To overcome this discrepancy, I scale
the mirror trade data to make them comparable to the trade flows from
WIOD. I set $\tilde{F}_{ij,t}^{1,\mathrm{ACD}}=\tilde{F}_{ij,t}^{\mathrm{ACD}}$
and $\tilde{F}_{ij,t}^{2,\mathrm{ACD}}=\tilde{F}_{ij,t}^{2}\cdot\tilde{F}_{ij,t}^{\mathrm{ACD}}/\tilde{F}_{ij,t}^{1}$,
for $\tilde{F}_{ij,t}^{\mathrm{ACD}}$ the noisy trade flow as used
in \citet{adao2017nonparametric}. There were also some trade flows
in the mirror trade dataset that reported zeros but had a large trade
flow in the WIOD. For these trade flows, I set the zero mirror trade
data entries equal to the positive WIOD entry. 

\subsection{\protect\label{subsec:Details_ACD_supplementary_analyses}Supplementary
Analyses}

\subsubsection{Winsorized Measurement Error Variances}

\begin{figure}[h]
\centering{}\includegraphics[scale=0.44]{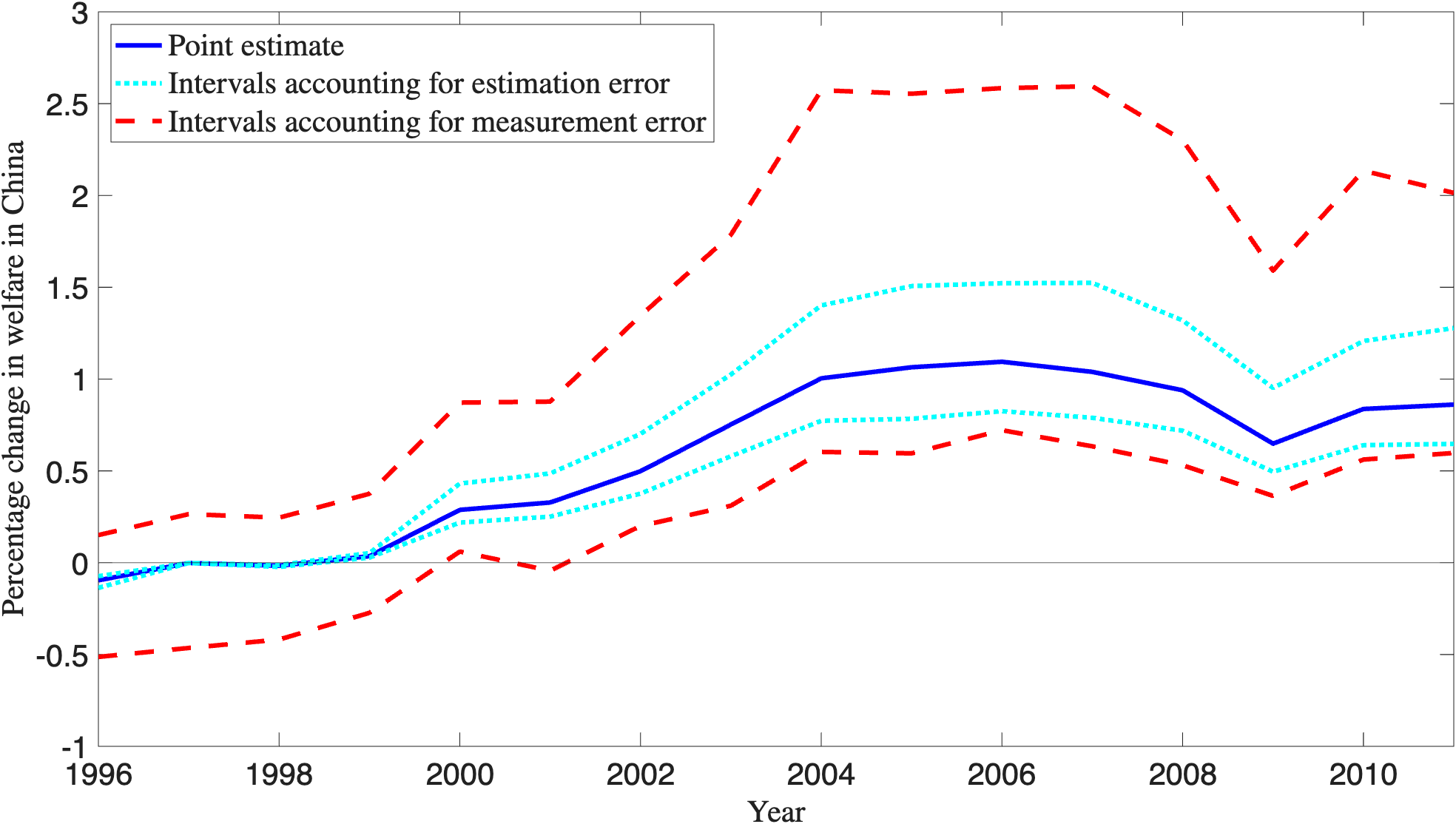}\caption{\protect\label{fig:ACD_winsor_0p2}Uncertainty quantification for
winsorized heteroskedastic normal shocks to $\left\{ \log F_{ij,t}\right\} $
for the percentage change in China's welfare due to the China shock.
The solid blue line is the estimate as reported in \citet{adao2017nonparametric},
the dotted light-blue lines denote the intervals accounting for estimation
error as reported in \citet{adao2017nonparametric}, and the dashed
red lines denote the intervals based on the estimated posterior distributions
$\pi^{\mathrm{post}}\left(g_{t}\left(\left\{ F_{ij,t}\right\} ,\varepsilon\right)|\left\{ \tilde{F}_{ij,t}\right\} ,\tilde{\vartheta}\right)$
for $t=1,...,T$.}
\end{figure}

The distribution of measurement error variances has a heavy right
tail, with the noisiest bilateral trade flow the one from Mexico to
Australia with a measurement error variance of 1.42. One might be
worried that this heavy tail drives the sensitivity to mismeasurement.
Figure \ref{fig:ACD_winsor_0p2} replicates Figure \ref{fig:ACD}
but now winsorizing the measurement error variances at 0.2, but keeping
the posterior variances constant. This amounts to winsorizing 27\%
of the trade flows. There are no substantial differences between Figures
\ref{fig:ACD_winsor_0p2} and \ref{fig:ACD}. 

\subsubsection{Accounting for Estimation Error and Measurement Error Simultaneously}

\begin{figure}[h]
\centering{}\includegraphics[scale=0.44]{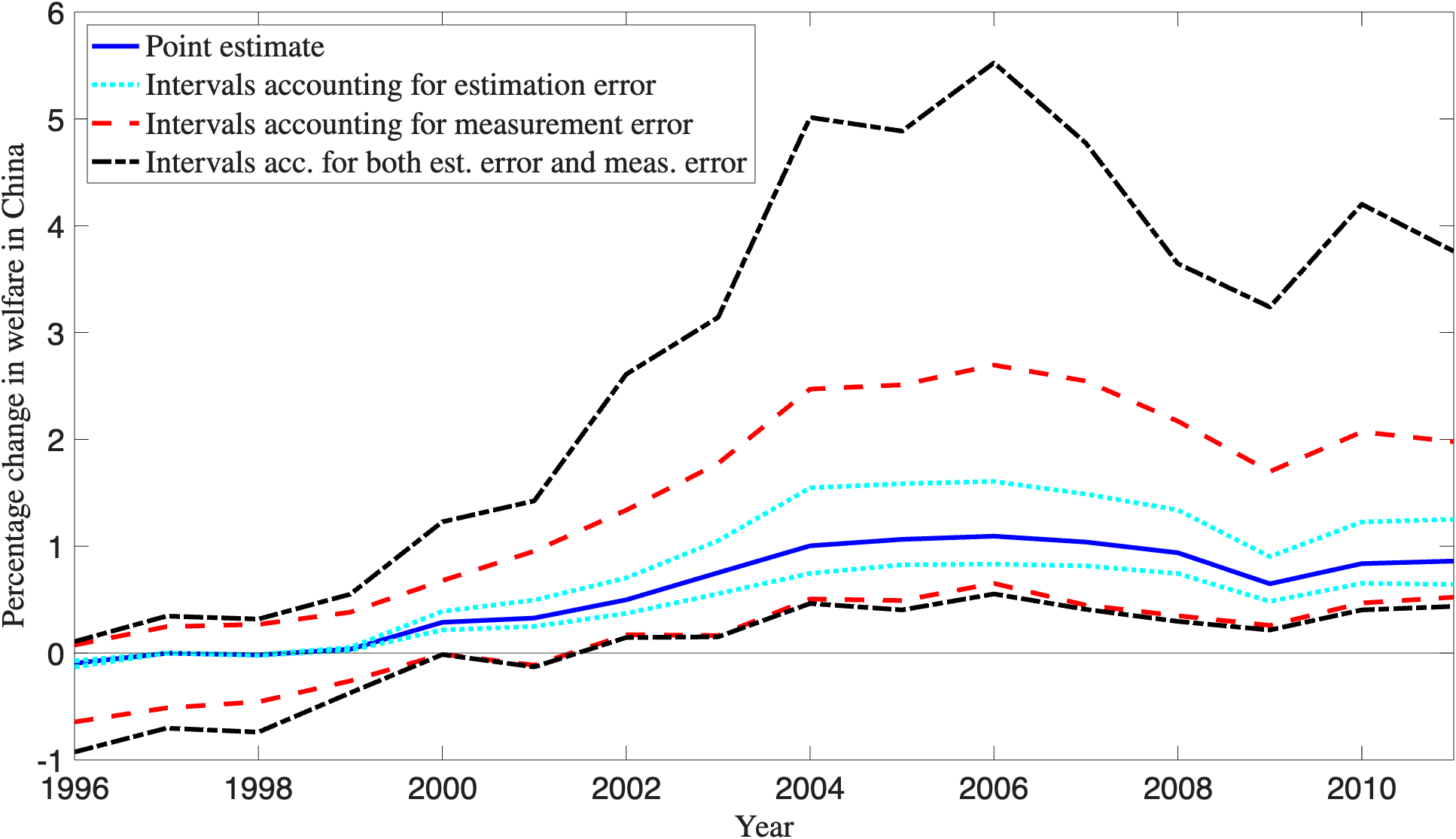}\caption{\protect\label{fig:ACD_eeme}Uncertainty quantification for heteroskedastic
normal shocks to $\left\{ \log F_{ij,t}\right\} $ for the percentage
change in China's welfare due to the China shock. The solid blue line
is the estimate as reported in \citet{adao2017nonparametric}, the
dotted light-blue lines denote the intervals accounting for estimation
error as reported in \citet{adao2017nonparametric}, the dashed red
lines denote the intervals based on the estimated posterior distributions
$\pi^{\mathrm{post}}\left(g_{t}\left(\left\{ F_{ij,t}\right\} ,\varepsilon\right)|\left\{ \tilde{F}_{ij,t}\right\} ,\tilde{\vartheta}\right)$
for $t=1,...,T$, and the dotted-dashed black lines denote the intervals
obtained using Algorithm \ref{alg:EB_UQ_k_ME_EE}.}
\end{figure}

As mentioned in Section \ref{subsec:ACD}, the authors of \citet{adao2017nonparametric}
estimate the trade elasticity $\varepsilon$ using GMM. Figure \ref{fig:ACD_eeme}
replicates Figure \ref{fig:ACD} but adds intervals obtained using
Algorithm \ref{alg:EB_UQ_k_ME_EE}. Interestingly, the upper bounds
increase substantially, while the lower bounds remain close to the
lower bounds of the intervals that account only for measurement error.

\subsubsection{Testing Normality Assumption and Gravity Model for the Prior}

\begin{figure}[h]
\centering{}\includegraphics[scale=0.44]{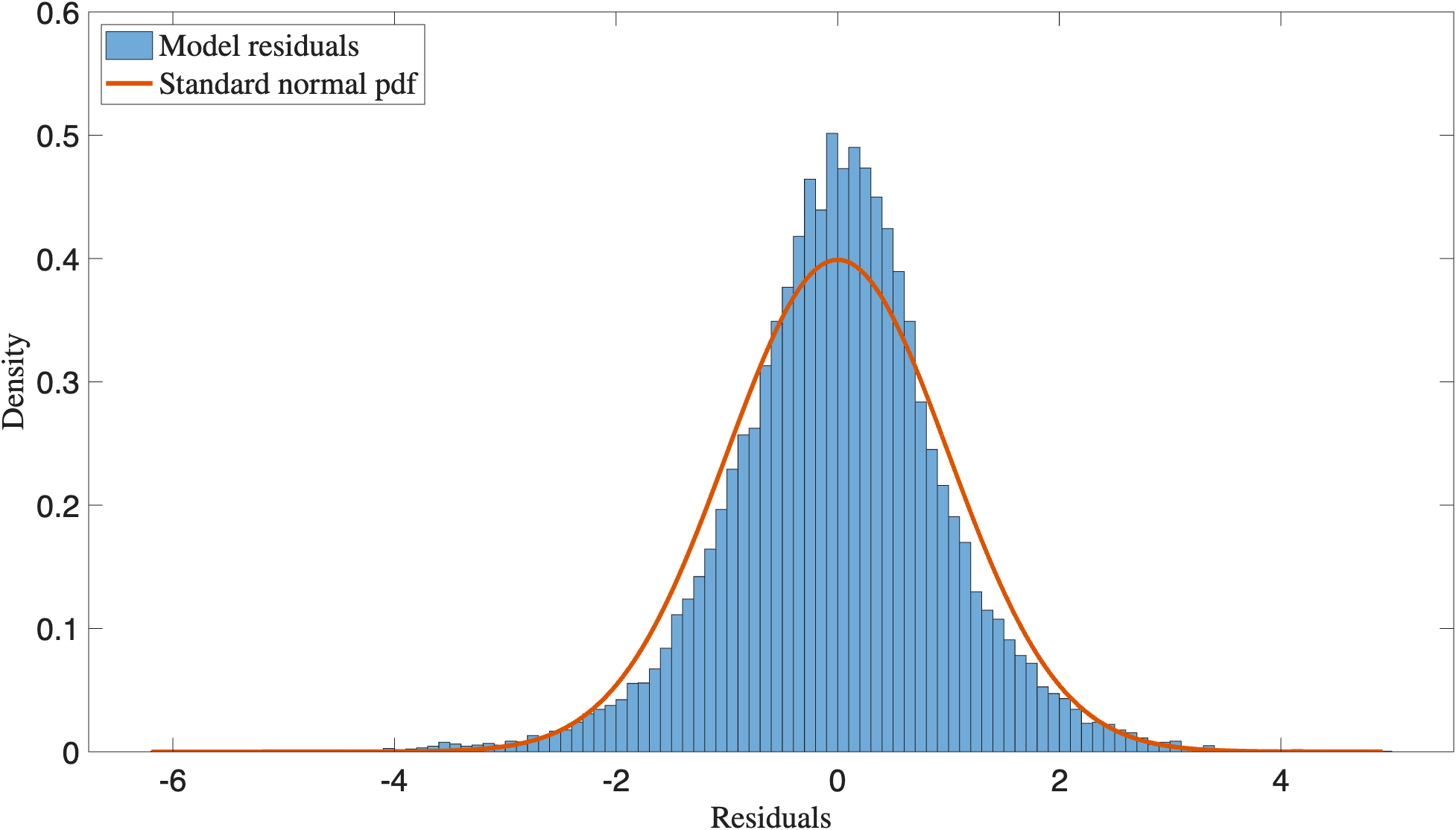}\caption{\protect\label{fig:ACD_normal_fit}Plot to compare the normalized
residuals with the probability density function of a standardized
normal distribution to check whether the normality assumption for
the prior is reasonable for \citet{adao2017nonparametric}.}
\end{figure}

As outlined in Remark \ref{rem:Checking_normality}, we can check
how reasonable the normality assumption is by comparing the histogram
of the normalized residuals with the probability density function
of a standardized normal distribution. The result can be found in
Figure \ref{fig:ACD_normal_fit}. It follows that the normality assumption
seems reasonable. 

\begin{figure}[h]
\centering{}\includegraphics[scale=0.44]{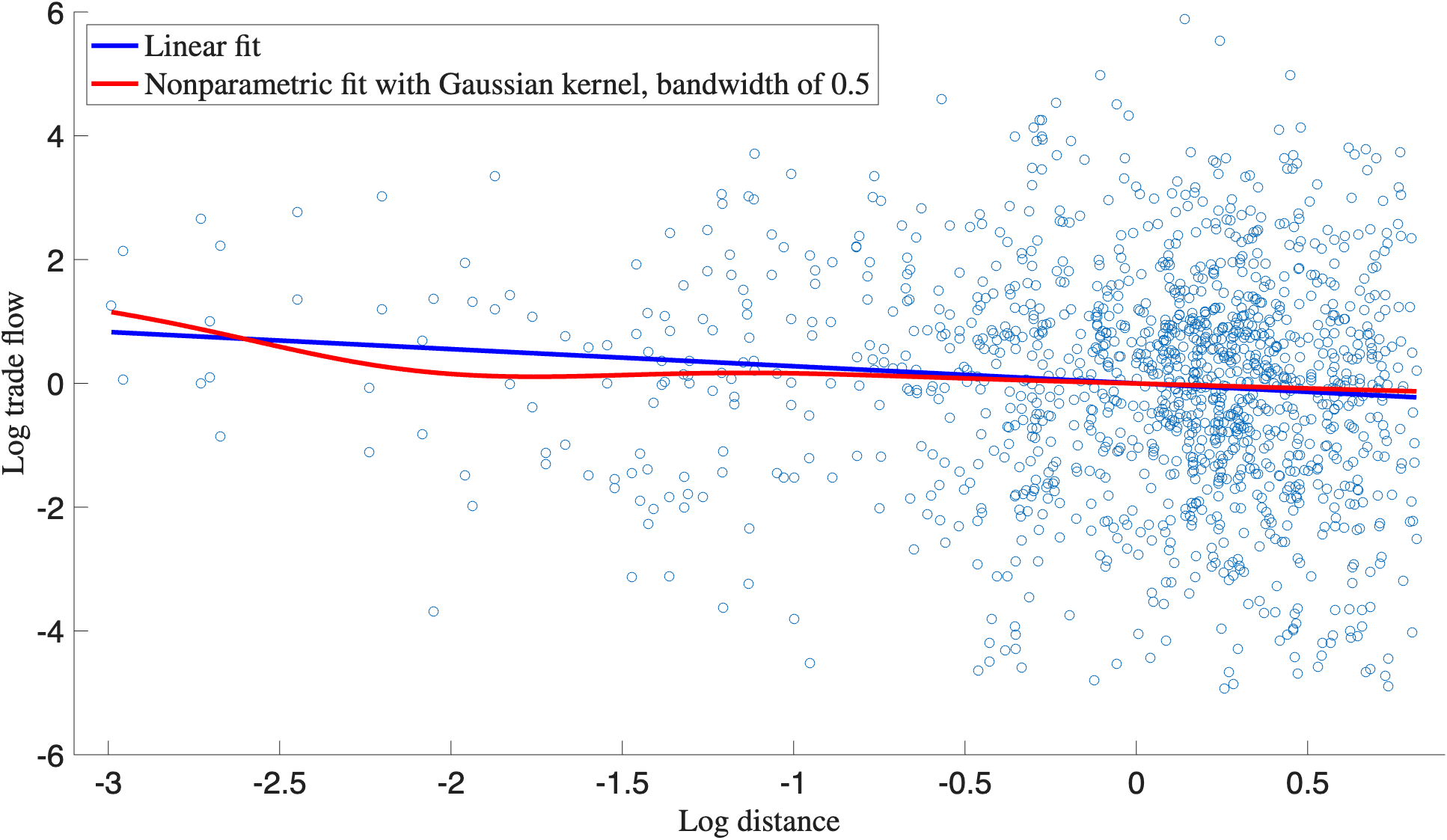}\caption{\protect\label{fig:ACD_grav_fit}Plot that follows \citet{allen201813}
to check whether the gravity model is reasonable for log trade flows
in 2011 from \citet{adao2017nonparametric}.}
\end{figure}

Concerning the gravity model, restricting attention to the year 2011,
the regression for the prior mean in Equation \eqref{eq:MLE_mirror_trade_reg}
has an adjusted R-squared of 0.95, and the coefficient on log distance
is -0.277 with a t-statistic of 3.346. Furthermore, Figure \ref{fig:ACD_grav_fit}
follows \citet{allen201813} by plotting a linear and nonparametric
fit of log trade flows against log distance, after partitioning out
the origin and destination fixed effects. Together, the high adjusted
R-squared and the good performance of the linear fit imply that the
gravity model is a reasonable choice for this setting. 

\section{\protect\label{sec:Details_AA}Details for Application \citet{allen2022welfare}}

\subsection{\protect\label{subsec:Details_AA_model}Model Details}

In the empirical application of \citet{allen2022welfare}, the authors
investigate what the returns on investment are of all the highway
segments of the US Interstate Highway network. Going forward, $\bar{L}$
denotes aggregate labor endowment, $\bar{Y}$ denotes total income
in the economy, $Q_{i}$ denotes the productivity of location $i$,
$A_{i}$ captures the level of amenities in location $i$, $\tau_{ij}$
denotes the travel cost between locations $i$ and $j$, $F_{ij}$
denotes the  traffic flow between locations $i$ and $j$, $y_{i}$
denotes total income of location $i$ as a share of the total income
in the economy, $\ell_{i}$ denotes the total labor in location $i$
as a share of the aggregate labor endowment, and $\chi$ captures
the (inverse of) the welfare of the economy. The parameter vector
is $\theta=\left(\alpha,\beta,\gamma,\nu\right)$, where $\alpha$
and $\beta$ control the strength of the productivity and amenity
externalities respectively, $\gamma$ is the shape parameter of the
Fréchet distributed idiosyncratic productivity shocks, and $\nu$
governs the strength of traffic congestion. 

It is shown in the paper that we can uniquely recover $\left(\left\{ y_{i}^{\mathrm{cf,prop}}\right\} ,\left\{ \ell_{i}^{\mathrm{cf,prop}}\right\} ,\chi^{\mathrm{cf,prop}}\right)$
given any change in the underlying infrastructure network $\left\{ \tau_{ij}^{\mathrm{cf,prop}}\right\} $
and baseline economic activity $\left\{ y_{i}\bar{Y}\right\} $, using
the system of equations 
\begin{align*}
 & \left(y_{i}^{\mathrm{cf,prop}}\right)^{\frac{1+\nu+\gamma}{1+\nu}}\left(\ell_{i}^{\mathrm{cf,prop}}\right)^{\frac{-\theta\left(1+\alpha+\nu\left(\alpha+\beta\right)\right)}{1+\nu}}\\
 & =\chi^{\mathrm{cf,prop}}\left(\frac{y_{i}\bar{Y}}{y_{i}\bar{Y}+\sum_{k}F_{ik}}\right)\left(y_{i}^{\mathrm{cf,prop}}\right)^{\frac{1+\nu+\gamma}{1+\nu}}\left(\ell_{i}^{\mathrm{cf,prop}}\right)^{\frac{\gamma\left(\beta-1\right)}{1+\nu}}\\
 & \qquad+\sum_{j}\left(\frac{F_{ij}}{y_{i}\bar{Y}+\sum_{k}F_{ik}}\right)\left(\tau_{ij}^{\mathrm{cf,prop}}\right)^{\frac{-\gamma}{1+\nu}}\left(y_{j}^{\mathrm{cf,prop}}\right)^{\frac{1+\gamma}{1+\nu}}\left(\ell_{j}^{\mathrm{cf,prop}}\right)^{\frac{-\gamma\left(1+\alpha\right)}{1+\nu}}\\
 & \left(y_{i}^{\mathrm{cf,prop}}\right)^{\frac{-\gamma+\nu}{1+\nu}}\left(\ell_{i}^{\mathrm{cf,prop}}\right)^{\frac{\gamma\left(1-\beta-\nu\left(\alpha+\beta\right)\right)}{1+\nu}}\\
 & =\chi^{\mathrm{cf,prop}}\left(\frac{y_{i}\bar{Y}}{y_{i}\bar{Y}+\sum_{k}F_{ki}}\right)\left(y_{i}^{\mathrm{cf,prop}}\right)^{\frac{-\gamma+\nu}{1+\nu}}\left(\ell_{i}^{\mathrm{cf,prop}}\right)^{\frac{\gamma\left(\alpha+1\right)}{1+\nu}}\\
 & \qquad+\sum_{j}\left(\frac{F_{ji}}{y_{i}\bar{Y}+\sum_{k}F_{ki}}\right)\left(\tau_{ij}^{\mathrm{cf,prop}}\right)^{\frac{-\gamma}{1+\nu}}\left(y_{j}^{\mathrm{cf,prop}}\right)^{\frac{-\gamma}{1+\nu}}\left(\ell_{j}^{\mathrm{cf,prop}}\right)^{\frac{\gamma\left(1-\beta\right)}{1+\nu}}.
\end{align*}
Having obtained $\chi^{\mathrm{cf,prop}}$, the proportional counterfactual
change in welfare is then calculated using
\[
W^{\mathrm{cf,prop}}=\frac{\left(\chi^{\mathrm{cf,prop}}\right)^{1/\gamma}}{\bar{L}^{\alpha+\beta}}.
\]

\subsection{\protect\label{subsec:Details_AA_calibration}Calibration Procedure
and Computational Details}

The default approach from Section \ref{sec:Default_approach} can
be applied. For the empirical Bayes step, the calibration of $\vartheta$,
we can use the baseline case with domain knowledge from Section \ref{subsec:Calibrating_vartheta_baseline}. 

When I run the code from \citet{allen2022welfare}, the returns of
investment for the links systematically differ slightly from the ones
in the paper. I scale my estimates so that the unperturbed estimates
align with the ones in the paper. 

\subsection{\protect\label{subsec:Details_AA_supplementary_analyses}Supplementary
Analyses}

\subsubsection{Probability that Rankings are Reversed}

We can learn more from the posterior distributions than just intervals.
It might be of interest what the expected probability is that the
ranking of the three links are reversed. The expected probability
that the ranking between link 1 and link 2 is reversed is 0.000, and
the expected probability that the ranking between link 2 and link
3 is 0.030.

\subsubsection{Testing Normality Assumption and Gravity Model for the Prior}

\begin{figure}[h]
\centering{}\includegraphics[scale=0.44]{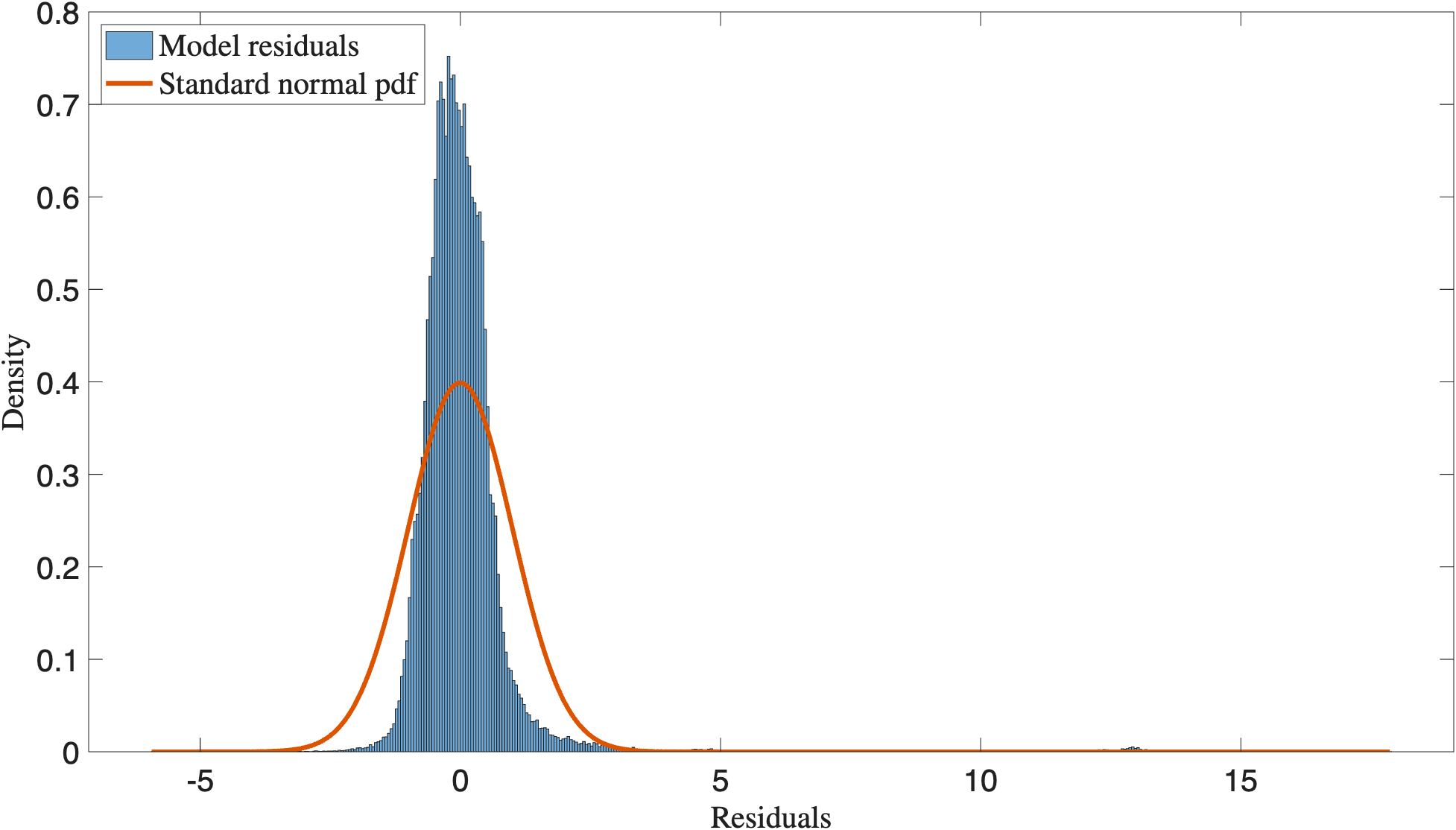}\caption{\protect\label{fig:AA_normal_fit}Plot to compare the normalized residuals
with the probability density function of a standardized normal distribution
to check whether the normality assumption for the prior is reasonable
for \citet{allen2022welfare}.}
\end{figure}

We can again check the reasonableness of the normality assumption
as per Remark \ref{rem:Checking_normality}. The result can be found
in Figure \ref{fig:AA_normal_fit}, and it follows that the normality
assumption is less reasonable compared to the setting of \citet{adao2017nonparametric}. 

\begin{figure}[h]
\centering{}\includegraphics[scale=0.44]{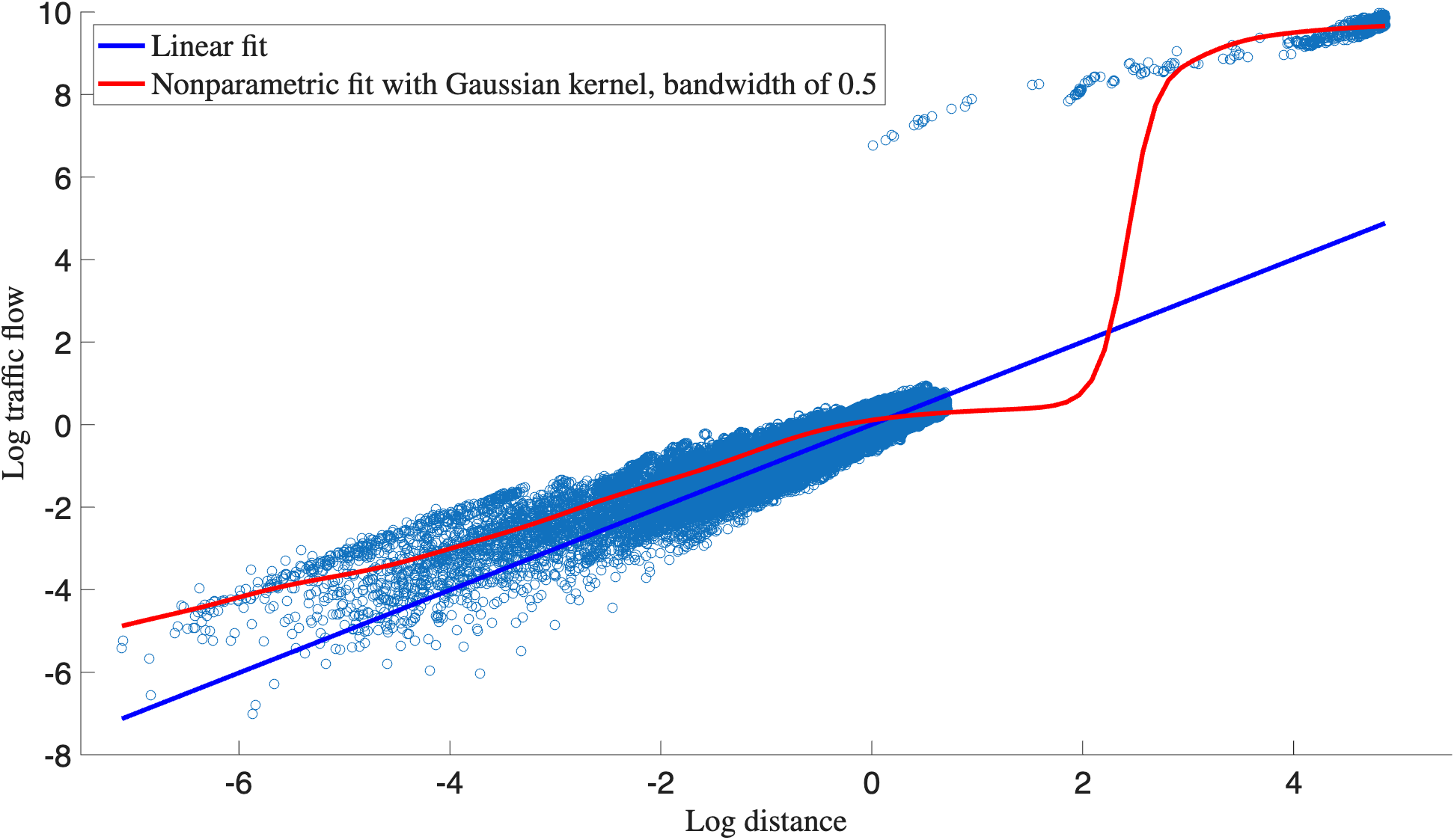}\caption{\protect\label{fig:AA_grav_fit}Plot that follows \citet{allen201813}
to check whether the gravity model is reasonable for log traffic flows
from \citet{allen2022welfare}.}
\end{figure}

Concerning the gravity model, the regression for the prior mean in
Equation \eqref{eq:prior_mean} has an adjusted R-squared of $0.9995$,
and the coefficient on log distance is $1.003$ with a t-statistic
of $1138$. It follows that log distance is an important driver of
log traffic flows, but not in a negative way as is common in gravity
models. Furthermore, Figure \ref{fig:AA_grav_fit} follows \citet{allen201813}
by plotting a linear and nonparametric fit of log traffic flows against
log distance, after partitioning out the origin and destination fixed
effects. Together, the high adjusted R-squared and the good performance
of the linear fit imply that the gravity model is a reasonable choice
for this setting. 

\section{\protect\label{sec:Misspecification_ME_prior}Misspecification of
the Measurement Error Model and Prior}

We are interested in the potential effects of misspecification of
the measurement error model or prior. Specifically, focusing on the
widely applicable default approach from Section \ref{sec:Default_approach},
we would like to know how the quantiles of the posterior distribution
of the counterfactual object of interest given the noisy flows change
when the assumptions of a normal measurement error model or a normal
prior do not hold. Suppose for exposition that there are no zeros
and the hyperparameters $\vartheta$ are known, so that we can obtain
the posterior distribution $\pi^{\mathrm{post}}\left(\gamma|\left\{ \log\tilde{F}_{ij}\right\} \right)$. 

\subsection{Measurement Error Model}

Let $L\left(\left\{ \log F_{ij}\right\} \right)=\pi^{\mathrm{me}}\left(\left\{ \log\tilde{F}_{ij}\right\} |\left\{ \log F_{ij}\right\} \right)$
denote the likelihood function of the noisy log flows $\left\{ \log\tilde{F}_{ij}\right\} $
given the true log flows $\left\{ \log F_{ij}\right\} $. For a given
$c\geq1$, define a density-ratio class of distributions to be the
set of all conditional distributions for $\left\{ \log\tilde{F}_{ij}\right\} $
with pdf $p$ such that
\[
p\in\mathcal{R}_{c}=\left\{ p\in P:\frac{1}{c}\cdot L\left(x\right)\leq p\left(x\right)\leq c\cdot L\left(x\right)\ \forall x\in\mathbb{R}^{n\left(n+1\right)}\right\} ,
\]
for $P$ the set of all pdfs. 

For uncertainty quantification, we are interested in the quantiles
of the posterior distribution $\pi^{\mathrm{post}}\left(h\left(\left\{ \log F_{ij}\right\} \right)|\left\{ \log\tilde{F}_{ij}\right\} \right)$
for a generic function $h\left(\cdot\right)$. Denote the $\alpha$-th
posterior quantile based on likelihood $p$ by $Q_{\pi,p,h}\left(\alpha\right)$.
\begin{prop}
\label{prop:quantiles}We have:
\begin{align*}
\underset{p\in\mathcal{R}_{c}}{\sup}Q_{\pi,p,h}\left(\alpha\right) & =Q_{\pi,L,h}\left(\frac{\alpha c^{2}}{1-\alpha+\alpha c^{2}}\right)\\
\underset{p\in\mathcal{R}_{c}}{\inf}Q_{\pi,p,h}\left(\alpha\right) & =Q_{\pi,L,h}\left(\frac{\alpha}{\alpha+\left(1-\alpha\right)c^{2}}\right).
\end{align*}
\end{prop}
So instead of reporting the interval
\[
\left[Q_{\pi,L,h}\left(\alpha/2\right),Q_{\pi,L,h}\left(1-\alpha/2\right)\right]
\]
one could report the robust interval
\[
\left[Q_{\pi,L,h}\left(\frac{\alpha}{\alpha+\left(2-\alpha\right)c^{2}}\right),Q_{\pi,L,h}\left(\frac{\left(2-\alpha\right)c^{2}}{\alpha+\left(2-\alpha\right)c^{2}}\right)\right].
\]
For example for $\alpha=0.05$ and $c=1.5$, we would consider the
$1.1\%$-quantile and the $98.9\%$-quantile, instead of the $2.5\%$-quantile
and the $97.5\%$-quantile, respectively. 

The result in Proposition \ref{prop:quantiles} follows from noting
that
\begin{align*}
 & \alpha=\int_{-\infty}^{q}\pi^{\mathrm{post}}\left(h\left(x\right)|\tilde{x}\right)dh\left(x\right)=\int_{x\in h^{-1}\left(\left[-\infty,q\right]\right)}\pi^{\mathrm{post}}\left(x|\tilde{x}\right)dx\\
\Rightarrow & \int_{x\in h^{-1}\left(\left[-\infty,q\right]\right)}p\left(x\right)\pi^{\mathrm{prior}}\left(x\right)dx=\frac{\alpha}{1-\alpha}\int_{x\not\in h^{-1}\left(\left[-\infty,q\right]\right)}p\left(x\right)\pi^{\mathrm{prior}}\left(x\right)dx.
\end{align*}
Focusing on the upper bound, it follows that we want to choose $p\left(x\right)$
on the left-hand side as small as possible and $p\left(x\right)$
on the right-hand side as large as possible for all $x$:
\begin{align*}
 & \frac{1}{c}\int_{x\in h^{-1}\left(\left[-\infty,q_{\mathrm{sup}}^{*}\right]\right)}L\left(x\right)\pi^{\mathrm{prior}}\left(x\right)dx=\frac{\alpha}{1-\alpha}c\int_{x\not\in h^{-1}\left(\left[-\infty,q_{\mathrm{sup}}^{*}\right]\right)}L\left(x\right)\pi^{\mathrm{prior}}\left(x\right)dx\\
\Rightarrow & \int_{-\infty}^{q_{\mathrm{sup}}^{*}}\pi^{\mathrm{post}}\left(h\left(x\right)|\left\{ \log\tilde{F}_{ij}\right\} \right)dh\left(x\right)=\frac{\alpha c^{2}}{1-\alpha+\alpha c^{2}}.
\end{align*}

\subsection{Prior}

Note that the likelihood $L$ and the prior $\pi^{\mathrm{prior}}$
enter the posterior in exactly the same way, so we can interpret the
procedure in the previous subsection also as sensitivity analysis
with respect to the prior. 

\section{\protect\label{sec:Calibration_mirror_trade_data}Calibration with
Mirror Trade Data}

\subsection{Model}

I use the mirror trade dataset from \citet{linsi2023problem}. This
dataset has two estimates of each bilateral trade flow, both as reported
by the exporter and as by the importer. I interpret this as observing
two independent noisy observations per time period for each bilateral
trade flow: $\left\{ \left\{ \tilde{F}_{ij,t}^{1},\tilde{F}_{ij,t}^{2}\right\} _{t=1}^{T}\right\} _{i\neq j}$.
It is helpful to rewrite the model:
\[
\begin{cases}
\begin{array}{c}
\mathrm{true\ zeros:}\\
\mathrm{spurious\ zeros:}\\
\mathrm{prior:}\\
\\\\\mathrm{likelihood:}\\
\\\\\\\\\end{array} & \begin{array}{c}
P_{ij,t}\sim\mathrm{Bern}\left(p_{ij}\right)\\
B_{ij,t}^{1},B_{ij,t}^{2}\sim\mathrm{Bern}\left(b_{ij}\right)\\
F_{ij,t}\sim P_{ij,t}\cdot\delta_{0}+\left(1-P_{ij,t}\right)\cdot e^{\mu_{ij,t}}\cdot e^{\eta_{ij,t}}\\
\mu_{ij,t}=\beta_{t}\log\mathrm{dist}_{ij}+\alpha_{i,t}^{\mathrm{orig}}+\alpha_{j,t}^{\mathrm{dest}}\\
\eta_{ij,t}\sim\mathcal{N}\left(0,s_{ij}^{2}\right)\\
\tilde{F}_{ij,t}^{1}|F_{ij,t}\sim\delta_{0}\cdot\mathbb{I}\left\{ F_{ij,t}=0\right\} \\
+\left[B_{ij,t}^{1}\cdot\delta_{0}+\left(1-B_{ij,t}^{1}\right)\cdot F_{ij,t}\cdot e^{\varepsilon_{ij,t}^{1}}\right]\cdot\mathbb{I}\left\{ F_{ij,t}>0\right\} \\
\tilde{F}_{ij,t}^{2}|F_{ij,t}\sim\delta_{0}\cdot\mathbb{I}\left\{ F_{ij,t}=0\right\} \\
+\left[B_{ij,t}^{2}\cdot\delta_{0}+\left(1-B_{ij,t}^{2}\right)\cdot F_{ij,t}\cdot e^{\varepsilon_{ij,t}^{2}}\right]\cdot\mathbb{I}\left\{ F_{ij,t}>0\right\} \\
\varepsilon_{ij,t}^{1},\varepsilon_{ij,t}^{2}\sim\mathcal{N}\left(0,\varsigma_{ij}^{2}\right)
\end{array}.\end{cases}
\]

\subsection{Bernoulli Parameters}

For a given bilateral trade flow from $i$ to $j$ in period $t$,
we can compute the ex-ante probability of observing a certain number
of zeros:
\begin{align*}
Pr\left\{ \mathrm{two\ observed\ zeros}\right\}  & =p_{ij}+\left(1-p_{ij}\right)\cdot b_{ij}^{2}\\
Pr\left\{ \mathrm{one\ observed\ zero}\right\}  & =2\cdot\left(1-p_{ij}\right)\cdot\left(1-b_{ij}\right)\cdot b_{ij}\\
Pr\left\{ \mathrm{no\ observed\ zeros}\right\}  & =\left(1-p_{ij}\right)\cdot\left(1-b_{ij}\right)^{2}.
\end{align*}
We can use the time variation to identify the probabilities on the
left-hand side:
\begin{align*}
\tilde{z}_{ij,2} & =\frac{1}{T}\sum_{t=1}^{T}\mathbb{I}\left\{ \tilde{F}_{ij,t}^{1}=0,\tilde{F}_{ij,t}^{2}=0\right\} \\
\tilde{z}_{ij,1} & =\frac{1}{T}\sum_{t=1}^{T}\mathbb{I}\left\{ \tilde{F}_{ij,t}^{1}=0,\tilde{F}_{ij,t}^{2}>0\mathrm{\ or\ }\tilde{F}_{ij,t}^{1}>0,\tilde{F}_{ij,t}^{2}=0\right\} \\
\tilde{z}_{ij,0} & =\frac{1}{T}\sum_{t=1}^{T}\mathbb{I}\left\{ \tilde{F}_{ij,t}^{1}>0,\tilde{F}_{ij,t}^{2}>0\right\} .
\end{align*}
A large $T$ is not required for identification, but it improves precision.
When $\tilde{z}_{ij,2},\tilde{z}_{ij,1},\tilde{z}_{ij,0}\in\left(0,1\right)$,
we can back out the estimated probability of a true zero $\tilde{p}_{ij}$
and the estimated probability of a spurious zero $\tilde{b}_{ij}$
by solving
\begin{align*}
\tilde{z}_{ij,2} & =\tilde{p}_{ij}+\left(1-\tilde{p}_{ij}\right)\cdot\tilde{b}_{ij}^{2}\\
\tilde{z}_{ij,1} & =2\cdot\left(1-\tilde{p}_{ij}\right)\cdot\left(1-\tilde{b}_{ij}\right)\cdot\tilde{b}_{ij}\\
\tilde{z}_{ij,0} & =\left(1-\tilde{p}_{ij}\right)\cdot\left(1-\tilde{b}_{ij}\right)^{2}.
\end{align*}
The solutions are
\[
\tilde{p}_{ij}=\max\left\{ 1-\frac{\left(\tilde{z}_{ij,1}+2\tilde{z}_{ij,0}\right)^{2}}{4\tilde{z}_{ij,0}},0\right\} ,\quad\tilde{b}_{ij}=\frac{\tilde{z}_{ij,1}}{\tilde{z}_{ij,1}+2\tilde{z}_{ij,0}}.
\]
I separately consider the possible cases where the estimated probabilities
$\left(\tilde{z}_{ij,2},\tilde{z}_{ij,1},\tilde{z}_{ij,0}\right)$
are not all strictly between 0 and 1:
\begin{enumerate}
\item $\tilde{z}_{ij,2}=1,\tilde{z}_{ij,1}=0,\tilde{z}_{ij,0}=0$: In this
case we observe only zeros so I set the estimated probability of a
true zero $\tilde{p}_{ij}$ to $1$, which makes the estimated probability
of a spurious zero $\tilde{b}_{ij}$ irrelevant.
\item $\tilde{z}_{ij,2}=0,\tilde{z}_{ij,1}=1,\tilde{z}_{ij,0}=0$: In this
case one country always reports a positive flow and the other reports
a zero flow. In this case I set the estimated probability of a true
zero $\tilde{p}_{ij}$ to $0$, and the estimated probability of a
spurious zero $\tilde{b}_{ij}$ to $0.5$.
\item $\tilde{z}_{ij,2}=0,\tilde{z}_{ij,1}=0,\tilde{z}_{ij,0}=1$: In this
case all reported flows are positive, so I set both the estimated
probability of a true zero $\tilde{p}_{ij}$ and the estimated probability
of a spurious zero $\tilde{b}_{ij}$ to $0$.
\item $\tilde{z}_{ij,2}\in\left(0,1\right),\tilde{z}_{ij,1}\in\left(0,1\right),\tilde{z}_{ij,0}=0$:
In this case there are no years with two reported positive flows.
In this case I set the estimated probability of a true zero $\tilde{p}_{ij}$
to $\tilde{z}_{ij,2}$, and the estimated probability of a spurious
zero $\tilde{b}_{ij}$ to $\tilde{z}_{ij,1}$.
\item $\tilde{z}_{ij,2}\in\left(0,1\right),\tilde{z}_{ij,1}=0,\tilde{z}_{ij,0}\in\left(0,1\right)$:
In this case some years have two zeros and other years have two positive
flows. In this case I set the estimated probability of a true zero
$\tilde{p}_{ij}$ to $\tilde{z}_{ij,2}$, and the estimated probability
of a spurious zero $\tilde{b}_{ij}$ to $0$.
\item $\tilde{z}_{ij,2}=0,\tilde{z}_{ij,1}\in\left(0,1\right),\tilde{z}_{ij,0}\in\left(0,1\right)$:
In this case there are no reported double zeros so I set the estimated
probability of a true zero $\tilde{p}_{ij}$ to $0$. I then solve
the system of equations:
\begin{align*}
\tilde{z}_{ij,1} & =\widetilde{Pr}\left\{ \mathrm{one\ observed\ zero}|\mathrm{no\ spurious\ zeros,\ observed\ zeros}<2\right\} \\
 & =\frac{2\tilde{b}_{ij}\left(1-\tilde{b}_{ij}\right)}{2\tilde{b}_{ij}\left(1-\tilde{b}_{ij}\right)+\left(1-\tilde{b}_{ij}\right)^{2}}=\frac{2\tilde{b}_{ij}}{1+\tilde{b}_{ij}}\\
\tilde{z}_{ij,0} & =\widetilde{Pr}\left\{ \mathrm{no\ observed\ zeros}|\mathrm{no\ spurious\ zeros,\ observed\ zeros}<2\right\} \\
 & =\frac{\left(1-\tilde{b}_{ij}\right)^{2}}{2\tilde{b}_{ij}\left(1-\tilde{b}_{ij}\right)+\left(1-\tilde{b}_{ij}\right)^{2}}=\frac{1-\tilde{b}_{ij}}{1+\tilde{b}_{ij}},
\end{align*}
and find
\[
\tilde{b}_{ij}=\frac{\tilde{z}_{ij,1}}{2-\tilde{z}_{ij,1}}.
\]
\end{enumerate}

\subsection{Measurement Error Variances}

We can combine the model equations to find:
\begin{align*}
\log\tilde{F}_{ij,t}^{1} & =\beta_{t}\log\mathrm{dist}_{ij}+\alpha_{i,t}^{\mathrm{orig}}+\alpha_{j,t}^{\mathrm{dest}}+\eta_{ij,t}+\varepsilon_{ij,t}^{1},\quad\tilde{F}_{ij,t}^{1}>0\\
\log\tilde{F}_{ij,t}^{2} & =\beta_{t}\log\mathrm{dist}_{ij}+\alpha_{i,t}^{\mathrm{orig}}+\alpha_{j,t}^{\mathrm{dest}}+\eta_{ij,t}+\varepsilon_{ij,t}^{2},\quad\tilde{F}_{ij,t}^{2}>0,
\end{align*}
for $i,j=1,...,n$ and $t=1,...,T$. Subtracting these two equations
yields
\[
\log\tilde{F}_{ij,t}^{1}-\log\tilde{F}_{ij,t}^{2}=\varepsilon_{ij,t}^{1}-\varepsilon_{ij,t}^{2}\sim\mathcal{N}\left(0,2\varsigma_{ij}^{2}\right),\quad\tilde{F}_{ij,t}^{1}>0,\tilde{F}_{ij,t}^{2}>0,
\]
 for $i,j=1,...,n$ and $t=1,...,T$. This suggests the estimator
\[
\tilde{\varsigma}_{ij}^{2}=\frac{\mathbb{I}\left\{ \sum_{t=1}^{T}\mathbb{I}\left\{ \tilde{F}_{ij,t}^{1}>0,\tilde{F}_{ij,t}^{2}>0\right\} >0\right\} }{\sum_{t=1}^{T}\mathbb{I}\left\{ \tilde{F}_{ij,t}^{1}>0,\tilde{F}_{ij,t}^{2}>0\right\} }\frac{1}{2}\sum_{t=1}^{T}\mathbb{I}\left\{ \tilde{F}_{ij,t}^{1}>0,\tilde{F}_{ij,t}^{2}>0\right\} \cdot\left(\log\tilde{F}_{ij,t}^{1}-\log\tilde{F}_{ij,t}^{2}\right)^{2}
\]
for $i,j=1,...,n$. So note that county-pairs with no entries with
two positive flows will have an estimated measurement error variance
of $0$. Note that the estimator is unbiased even with access to one
period of mirror trade data (assuming both flows are non-negative).
Obtaining estimators for the measurement error variances is what \citet{walters2024empirical}
calls the estimation step.

\subsection{Prior Means}

For the calibration of $\left(\left\{ \beta_{t}\right\} ,\left\{ \alpha_{i,t}^{\mathrm{exp}}\right\} ,\left\{ \alpha_{j,t}^{\mathrm{imp}}\right\} \right)$,
I use $\tilde{F}_{ij,t}=\tilde{F}_{ij,t}^{1}$. We then know that
\begin{equation}
\log\tilde{F}_{ij,t}\sim\mathcal{N}\left(\beta_{t}\log\mathrm{dist}_{ij}+\alpha_{i,t}^{\mathrm{orig}}+\alpha_{j,t}^{\mathrm{dest}},s_{ij}^{2}+\varsigma_{ij}^{2}\right),\quad\mathrm{for}\ \tilde{F}_{ij,t}>0,\label{eq:MLE_mirror_trade}
\end{equation}
for $i,j=1,...,n$ and $t=1,...,T$.\footnote{Alternatively, one could use the average of the two measurements $0.5\left(\tilde{F}_{ij,t}^{1}+\tilde{F}_{ij,t}^{2}\right)$.
In this case Equation \eqref{eq:MLE_mirror_trade} changes to $\log\tilde{F}_{ij,t}\sim\mathcal{N}\left(\beta_{t}\log\mathrm{dist}_{ij}+\alpha_{i,t}^{\mathrm{orig}}+\alpha_{j,t}^{\mathrm{dest}},s_{ij}^{2}+0.5\cdot\varsigma_{ij}^{2}\right)$. } Using maximum likelihood estimation, it follows that the prior mean
parameters can be estimated from the \textit{within-period} regressions
\begin{equation}
\log\tilde{F}_{ij,t}=\beta_{t}\log\mathrm{dist}_{ij}+\alpha_{i,t}^{\mathrm{orig}}+\alpha_{j,t}^{\mathrm{dest}}+\zeta_{ij,t},\quad\mathrm{for}\ \tilde{F}_{ij,t}>0,\label{eq:MLE_mirror_trade_reg}
\end{equation}
for $t=1,...,T$, with $\zeta_{ij,t}$ an error term. The estimated
prior means are
\begin{align*}
\tilde{\mu}_{ij,t} & =\left(\tilde{\beta}_{t}\log\mathrm{dist}_{ij}+\tilde{\alpha}_{i,t}^{\mathrm{orig}}+\tilde{\alpha}_{j,t}^{\mathrm{dest}}\right)\cdot\mathbb{I}\left\{ \tilde{F}_{ij,t}>0\right\} \\
 & \quad+\frac{\mathbb{I}\left\{ \sum_{s=1}^{T}\mathbb{I}\left\{ \tilde{F}_{ij,s}>0\right\} >0\right\} }{\sum_{s=1}^{T}\mathbb{I}\left\{ \tilde{F}_{ij,s}>0\right\} }\cdot\sum_{s=1}^{T}\left\{ \tilde{\beta}_{s}\log\mathrm{dist}_{ij}+\tilde{\alpha}_{i,s}^{\mathrm{orig}}+\tilde{\alpha}_{j,s}^{\mathrm{dest}}\right\} ,
\end{align*}
for $i,j=1,...,n$ and $t=1,...,T$. Note that for zero flows, the
prior mean is imputed using an \textit{across-period} average, and
$\tilde{\mu}_{ij,t}$ is only zero if $\tilde{F}_{ij,t}$ is zero
in all time periods for that country pair. 

\subsection{Prior Variances}

From Equation \eqref{eq:MLE_mirror_trade} it follows that the posterior
variances can be estimated by
\begin{align*}
\tilde{s}_{ij}^{2} & =\max\left\{ \widetilde{\Var}\left(\log\tilde{F}_{ij,t}-\tilde{\mu}_{ij,t}|\tilde{F}_{ij,t}>0\right)-\tilde{\varsigma}_{ij}^{2},0\right\} ,
\end{align*}
for $i,j=1,...,n$. Here, I again impute across periods for zero flows.
Obtaining estimators for the prior means and variances is what \citet{walters2024empirical}
calls the deconvolution step.

\subsection{Shrinking Variance Estimates}

To leverage country information and the fact that importers and exporters
can differ in their reliability, and reduce the variability for $\left\{ \tilde{\varsigma}_{ij}^{2}\right\} $
and $\left\{ \tilde{s}_{ij}^{2}\right\} $, I fit the models
\begin{equation}
\tilde{\varsigma}_{ij}^{2}=e^{\kappa_{i}^{\varsigma,\mathrm{orig}}+\kappa_{j}^{\varsigma,\mathrm{dest}}+u_{ij}^{\varsigma}}\quad\mathrm{and}\quad\tilde{s}_{ij}^{2}=e^{\kappa_{i}^{s,\mathrm{orig}}+\kappa_{j}^{s,\mathrm{dest}}+u_{ij}^{s}},\label{eq:fixedEffects_exp_imp-1}
\end{equation}
for $i,j=1,...,n$, with $\kappa_{i}^{\varsigma,\mathrm{orig}}$,
$\kappa_{j}^{\varsigma,\mathrm{dest}}$, $\kappa_{i}^{s,\mathrm{orig}}$
and $\kappa_{j}^{s,\mathrm{dest}}$ country-origin and country-destination
fixed effects and $u_{ij}^{\varsigma}$ and $u_{ij}^{s}$ error terms.
Then, rather than using $\tilde{\varsigma}_{ij}^{2}$ and $\tilde{s}_{ij}^{2}$
I will use the fitted values $\mathring{\varsigma}_{ij}^{2}=e^{\tilde{\kappa}_{i}^{\varsigma,\mathrm{orig}}+\tilde{\kappa}_{j}^{\varsigma,\mathrm{dest}}}$
and $\mathring{s}_{ij}^{2}=e^{\tilde{\kappa}_{i}^{s,\mathrm{orig}}+\tilde{\kappa}_{j}^{s,\mathrm{dest}}}$. 

\subsection{Posterior Draws}

It follows that the estimated posterior distribution for the true
flow between location $i$ and $j$, $F_{ij,t}$, given its noisy
version, $\tilde{F}_{ij,t}$ is given by
\begin{equation}
F_{ij,t}|\tilde{F}_{ij,t},\tilde{\vartheta}\sim\begin{cases}
\begin{array}{c}
Q_{ij}\cdot\delta_{0}+\left(1-Q_{ij}\right)\cdot e^{\mathcal{N}\left(\tilde{\mu}_{ij,t},\mathring{s}_{ij}^{2}\right)}\\
\exp\left\{ \mathcal{N}\left(\frac{\mathring{s}_{ij}^{2}}{\mathring{s}_{ij}^{2}+\mathring{\varsigma}_{ij}^{2}}\log\tilde{F}_{ij,t}+\frac{\mathring{\varsigma}_{ij}^{2}}{\mathring{s}_{ij}^{2}+\mathring{\varsigma}_{ij}^{2}}\tilde{\mu}_{ij,t},\left(\frac{1}{\mathring{s}_{ij}^{2}}+\frac{1}{\mathring{\varsigma}_{ij}^{2}}\right)^{-1}\right)\right\} 
\end{array} & \begin{array}{c}
\tilde{F}_{ij}=0\\
\tilde{F}_{ij}>0
\end{array},\end{cases}\label{eq:Posterior_mirror_trade}
\end{equation}
for $i,j=1,...,n$ and $t=1,...,T$, where $Q_{ij}\sim\mathrm{Bern}\left(\frac{\tilde{p}_{ij}}{\tilde{p}_{ij}+\tilde{b}_{ij}\left(1-\tilde{p}_{ij}\right)}\right)$. 

\subsection{Diagnostics}

From Equation \eqref{eq:MLE_mirror_trade}, one can verify how reasonable
the normality assumption on the prior and measurement error model
is by comparing the histogram of the normalized residuals 
\[
\left\{ \frac{\log\tilde{F}_{ij,t}-\tilde{\mu}_{ij,t}}{\sqrt{\mathring{s}_{ij}^{2}+\mathring{\varsigma}_{ij}^{2}}}\right\} _{i,j,t,\ \tilde{F}_{ij,t}>0}
\]
with the probability density function of a standard normal distribution.
To further check the reasonableness of the gravity prior, we can look
at the adjusted R-squared of the gravity regressions in Equation \eqref{eq:MLE_mirror_trade_reg},
and, following \citet{allen201813}, plot the log flows against the
log distance, after partitioning out the origin and destination fixed
effects.

\subsection{Computational Implementation Details}

In the case where for all years one country reports only positive
flows and the other country reports only NAs, I replace the NAs by
the positive flows. After this initial replacement step, I replace
the remaining NAs by zeros.

\section{Details for Armington Model}

\subsection{\protect\label{subsec:Derivation_Armington}Derivation of System
of Equations for $\left\{ Y_{i}^{\mathrm{prop}}\right\} $}

Rearranging Equation \eqref{eq:gravity} and recalling that $\lambda_{ij}=F_{ij}/E_{j}$
yields:
\begin{equation}
\lambda_{ij}=\frac{\left(\tau_{ij}Y_{i}\right)^{-\varepsilon}\chi_{ij}}{\sum_{k}\left(\tau_{kj}Y_{k}\right)^{-\varepsilon}\chi_{kj}},\qquad i,j=1,...,n.\label{eq:lambda}
\end{equation}
Next, plugging in Equations \eqref{eq:trade_deficit} and \eqref{eq:lambda}
into Equation \eqref{eq:gravity} yields
\[
F_{ij}=\lambda_{ij}\left(1+\kappa_{j}\right)Y_{j},\qquad i,j=1,...,n.
\]
If we sum over $j$, we can use $Y_{i}=\sum_{\ell=1}^{n}F_{i\ell}$
to find
\begin{equation}
Y_{i}=\sum_{j=1}^{n}\lambda_{ij}\left(1+\kappa_{j}\right)Y_{j},\qquad i=1,...,n.\label{eq:Y}
\end{equation}
In the counterfactual equilibrium, Equation \eqref{eq:Y} should still
hold. Because $\kappa_{i}$ is constant across equilibria for all
$i$, this results in:
\begin{equation}
Y_{i}^{\mathrm{cf,prop}}Y_{i}=\sum_{j=1}^{n}\lambda_{ij}^{\mathrm{cf,prop}}\lambda_{ij}\left(1+\kappa_{j}\right)Y_{j}^{\mathrm{cf,prop}}Y_{j},\qquad i=1,...,n.\label{eq:Y_prop}
\end{equation}
Similarly, Equation \eqref{eq:lambda} should still hold in equilibrium.
Using that $\chi_{ij}$ is constant across equilibria for all $i,j$,
we find
\begin{align}
\lambda_{ij}^{\mathrm{cf,prop}} & =\frac{1}{\lambda_{ij}}\frac{\left(\tau_{ij}^{\mathrm{cf,prop}}\tau_{ij}Y_{i}^{\mathrm{cf,prop}}Y_{i}\right)^{-\varepsilon}\chi_{ij}}{\sum_{k}\left(\tau_{kj}^{\mathrm{cf,prop}}\tau_{kj}Y_{k}^{\mathrm{cf,prop}}Y_{k}\right)^{-\varepsilon}\chi_{kj}}\nonumber \\
 & =\frac{1}{\lambda_{ij}}\frac{\left(\tau_{ij}^{\mathrm{cf,prop}}Y_{i}^{\mathrm{cf,prop}}\right)^{-\varepsilon}\frac{\left(\tau_{ij}Y_{i}^{\mathrm{}}\right)^{-\varepsilon}\chi_{ij}}{\sum_{\ell}\left(\tau_{\ell j}Y_{\ell}\right)^{-\varepsilon}\chi_{\ell j}}}{\sum_{k}\left(\tau_{kj}^{\mathrm{cf,prop}}Y_{k}^{\mathrm{cf,prop}}\right)^{-\varepsilon}\frac{\left(\tau_{kj}Y_{k}\right)^{-\varepsilon}\chi_{kj}}{\sum_{\ell}\left(\tau_{\ell j}Y_{\ell}\right)^{-\varepsilon}\chi_{\ell j}}}\nonumber \\
 & =\frac{\left(\tau_{ij}^{\mathrm{cf,prop}}Y_{i}^{\mathrm{cf,prop}}\right)^{-\varepsilon}}{\sum_{k}\lambda_{kj}\left(\tau_{kj}^{\mathrm{cf,prop}}Y_{k}^{\mathrm{cf,prop}}\right)^{-\varepsilon}},\qquad i,j=1,...,n.\label{eq:lambda_prop}
\end{align}
Finally, combining Equations \eqref{eq:Y_prop} and \eqref{eq:lambda_prop}
yields the desired expression
\[
Y_{i}^{\mathrm{cf,prop}}Y_{i}=\sum_{j}\frac{\left(\tau_{ij}^{\mathrm{cf,prop}}Y_{i}^{\mathrm{cf,prop}}\right)^{-\varepsilon}}{\sum_{k}\lambda_{kj}\left(\tau_{kj}^{\mathrm{cf,prop}}Y_{k}^{\mathrm{cf,prop}}\right)^{-\varepsilon}}\lambda_{ij}\left(1+\kappa_{j}\right)Y_{j}^{\mathrm{cf,prop}}Y_{j},\qquad i=1,...,n.
\]

\subsection{\protect\label{subsec:Other_countries}Results for Other Countries}

\begin{sidewaysfigure}[ph]
\centering{}\includegraphics[scale=0.7]{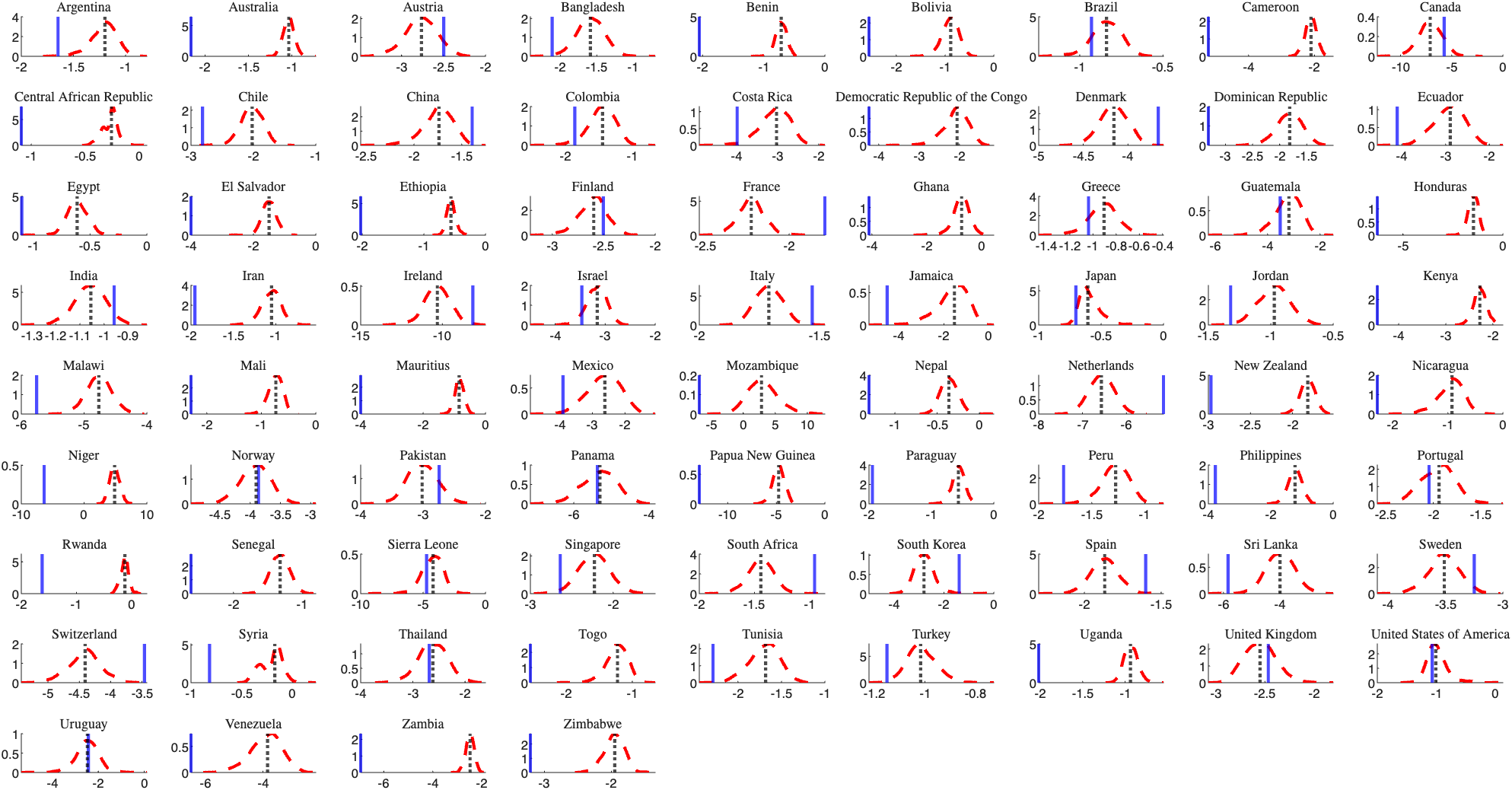}\caption{\protect\label{fig:Arm_gamma_ME_all_countries}Uncertainty quantification
for the Armington model. The counterfactual object of interest is
the percentage change in welfare (real consumption) after a 10\% increase
in all bilateral trade costs. The solid blue line denotes the point
estimate $g\left(\left\{ \tilde{F}_{ij}\right\} ,5\right)$, the dashed
red line denotes the smoothed estimated posterior distribution $\pi^{\mathrm{post}}\left(g\left(\left\{ F_{ij}\right\} ,5\right)|\left\{ \tilde{F}_{ij}\right\} ,\tilde{\vartheta}\right)$,
and the dotted black line denotes the median of $\pi^{\mathrm{post}}\left(g\left(\left\{ F_{ij}\right\} ,5\right)|\left\{ \tilde{F}_{ij}\right\} ,\tilde{\vartheta}\right)$.}
\end{sidewaysfigure}

Figure \ref{fig:Arm_gamma_ME_all_countries} reproduces Figure \ref{fig:Arm_gamma_ME}
for all 76 countries in the sample.

\subsection{\protect\label{subsec:Details_Armington}Calibration Procedure and
Computational Details}

The default approach from Section \ref{sec:Default_approach} can
be applied. For the empirical Bayes step, the calibration of $\vartheta$,
we can use the mirror trade data setting from Section \ref{subsec:Calibrating_vartheta_mirror_trade}.
To construct $\left\{ F_{ij}\right\} $, I use the mirror trade data
for bilateral flows $\left\{ F_{ij}\right\} _{i\neq j}$ and the trade
flow data from \citet{waugh2010international} for own-country flows
$\left\{ F_{ii}\right\} $. Because the mirror trade data report zero
bilateral trade flows for Belgium, I exclude it from the analysis,
resulting in a sample of 76 countries. 

\section{Appendix Figures}


\clearpage
\end{document}